\documentclass[a4paper,fleqn,usenatbib]{mnras} %,draft

\usepackage{multirow}
\usepackage{booktabs}
\usepackage[flushleft]{threeparttable}
\usepackage{xcolor}
\usepackage{booktabs,caption,fixltx2e}

\usepackage[final]{changes}
\makeatletter
\@namedef{Changes@AuthorColor}{black}
\colorlet{Changes@Color}{black}
\makeatother

\usepackage[T1]{fontenc}
\usepackage{ae,aecompl}

\usepackage{graphicx}	% Including figure files
\usepackage{amsmath}	% Advanced maths commands
\usepackage{amssymb}	% Extra maths symbols
\usepackage{graphics}
\usepackage{color}
\usepackage{textcomp}     % for \textquotesingle macro

\usepackage{natbib}
\usepackage{pdflscape}
\usepackage[export]{adjustbox}% http://ctan.org/pkg/adjustbox

\newcommand{\etal}{et~al.\ }
\newcommand{\Ha}{H$\alpha$ }
\newcommand{\hi}{H{\sc i}}

\def\kms{km~s$^{-1}$}

\def\msun{M$_\odot$}
\def\lsun{L$_\odot$}
\def\mhi{M$_\textrm{\hi}$}
\def\H2{H$_2$}
\def\Mst{M$_*$}
\def\wone{\textsc{w1}}
\def\wtwo{\textsc{w2}}
\def\wthree{\textsc{w3}}
\def\wfour{\textsc{w4}}

\title[Gas-rich low mass central galaxies]{xGASS: Gas-rich
  central galaxies in small groups and their connections to
  cosmic web gas feeding }

\author[Janowiecki et al.]{
Steven~Janowiecki,$^{1}$\thanks{E-mail: steven.janowiecki@icrar.org (SJ)}
Barbara~Catinella,$^{1}$
Luca~Cortese,$^{1}$
Am\'elie Saintonge,$^{2}$\newauthor
Toby Brown,$^{1,3}$
Jing Wang$^{4}$\\
\scriptsize
$^{1}$International Center for Radio Astronomy Research (ICRAR), M468,
The University of Western Australia, 35 Stirling Highway,  Crawley,
WA, 6009, Australia\\
\scriptsize
$^{2}$Department of Physics and Astronomy, University College London,
Gower Place, London WC1E 6BT, UK, \\
\scriptsize
$^{3}$Centre for Astrophysics and Supercomputing, Swinburne University
of Technology, Hawthorn, VIC 3122, Australia, \\
\scriptsize
$^{4}$CSIRO Astronomy \& Space Science, Australia Telescope National
Facility, P.O. Box 76, Epping, NSW 1710, Australia \\
}

\date{Accepted to MNRAS}
\pubyear{2016}

\begin{document}
\label{firstpage}
\pagerange{\pageref{firstpage}--\pageref{lastpage}}
\maketitle

% Abstract of the paper
\begin{abstract}

We use deep \hi \, observations obtained as part of the extended
\textit{GALEX}
Arecibo SDSS survey (xGASS) to study the cold gas properties of central
galaxies across environments. We find that, below stellar masses of
$10^{10.2}$~\msun, central galaxies in groups have an average atomic
hydrogen {gas fraction} $\sim$0.3dex higher than those in isolation at the
same stellar mass. At these {stellar}
masses, group central galaxies are {usually found in small
  groups of N=2 members}. The higher \hi \, content in these low mass group
central 
galaxies is mirrored by their higher average star formation activity
and molecular hydrogen content. At larger stellar masses,
this difference disappears and central galaxies in groups have similar
(or even smaller) gas 
reservoirs and star formation activity compared to those in isolation.
We discuss possible scenarios able to explain our findings and suggest
that the higher gas content in low mass group central galaxies is
likely due to contributions from the cosmic web or \hi-rich minor
mergers, which also fuel their enhanced star formation activity. 

%pdfendlink nesting issue:
%\vspace{1cm}

%Abstract (<=250 words)
\end{abstract}

% Select between one and six entries from the list of approved keywords.
% Don't make up new ones.
%\begin{keywords}
%keyword1 -- keyword2 -- keyword3
%\end{keywords}

%%%%%%%%%%%%%%%%%%%%%%%%%%%%%%%%%%%%%%%%%%%%%%%%%%
%%%%%%%%%%%%%%%%% BODY OF PAPER %%%%%%%%%%%%%%%%%%

\section{Introduction}

Studies have long shown a relationship between
galaxy morphology  and environmental density \citep[e.g.,][]{hubble31,
  dressler80, postmangeller84}. At high densities, galaxy
clusters are predominantly inhabited by gas-poor, red, passive
galaxies, while increasingly low density areas are populated by
galaxies which are increasingly blue, gas-rich, and actively
star-forming. A strong relation has been shown between a galaxy's
morphology and its cluster-centric radius \citep{whitmore93}, which
demonstrates the connections between environmental density and galaxy
transformations. Galaxies falling into rich clusters are observed to
experience rapid evolutionary transformations through dramatic
\added{mechanisms including ram-pressure stripping \citep{chung09} and
  starbursts \citep[see also][]{boselli06}.}
%mechanisms \citep[e.g., ram-pressure stripping, starbursts; see
%  also][]{boselli06}.

While striking and dramatic, these rapid transformations in
high-density environments are not the most important environmental
mechanism of galaxy evolution. Studies have shown that cluster infall
alone is insufficient to process field galaxies into cluster galaxies
while still maintaining observed scaling relations across environments
\citep{blantonmoustakas2009}. In order to maintain both global scaling
relations and the morphology-density relation, galaxies must
experience significant evolution through pre-processing in small
groups before they eventually merge into larger clusters. This
pre-processing can occur via mergers \citep{mihos04}, through gas
interactions \citep{fujita04}, or through tidal interactions
\citep{moore98}, and has been observed in galaxy groups in the local
Universe \citep{cortese06}. %,chung09}. 

Even though pre-processing makes a significant contribution to galaxy
evolution, it is difficult to study in small groups. First, 
galaxy groups (with $\lesssim$10 members) are 
difficult to consistently identify in optical galaxy surveys for
statistical reasons
\citep[see Section~\ref{sec:env} and][]{berlind06}. Second,
incompleteness in optically selected 
group catalogs is especially problematic for small groups, whose 
satellite members are often too faint for optical spectroscopy, but
can be identified by {deep \hi \, observations \citep{kern08}
  and blind \hi \, surveys \citep[e.g.,][]{hess13,odekon16}}. Third,
since 
gas-removal is one of the hallmarks of group pre-processing, the
most-processed galaxies will also be the most difficult (and
important) to detect in \hi \, and \H2.

These challenges have lead to a wide variety of results in the
literature. In recent optical studies of environment,
\citet{bamford08} found that at a fixed optical colour, the
morphology-density relationship disappears, while \citet{park07} found
that at a fixed morphology and stellar mass, no trends with
environmental density are observed (in colour, concentration, size,
star formation rate, etc.). Different studies have found that a
galaxy's host dark matter halo mass is the primary driver behind
environmental effects \citep[e.g.,][]{blantonberlind07} while others
conclude that the local density field drives environmental effects
\citep[e.g.,][]{kauffmann04}.

\hi \, studies of otherwise similar galaxies across different environments
have demonstrated that \hi-deficient galaxies are common in the
high-density cluster environment \citep{giovanellihaynes85, solanes01}
and also in the lower density group environments
\citep{verdes-montenegro2001, kilborn09}. However, observations have
also shown that \hi-rich galaxies in groups are more likely to be found
in \hi-rich environments \citep{wang15}, analogous to the conformity of
galaxy colours in groups and clusters found by
\citet{kauffmann10}. Continuing to the smallest group scales,
simulations and observations of galaxies in pairs have
found that they are enhanced in \hi \, \citep{tonnesen12} and SFR
\citep{lambas03, patton13} compared to un-paired galaxies.

Taken together, most \hi \, studies of
environment comprise a heterogeneous set of observations with
a variety of sensitivities, sample selections, and multi-wavelength
coverage. Blind \hi \, surveys such as the Arecibo Legacy Fast ALFA
\citep[ALFALFA,][]{giovanelli05,haynes11} survey are providing large samples of
galaxies, but cannot observe the gas-poor regime (i.e., those in
group or cluster environments) except for the most nearby galaxies
\citep{gavazzi13}.

The gas-rich population of ALFALFA galaxies has been used by
\citet{hess13} to study a sample of galaxy groups. They find that the
fraction of \hi-detected group members decreases as group membership
increases. ALFALFA \hi \, data have also been used in stacking 
analyses \citep[e.g.,][]{fabello11}, which combine \hi \, spectra from
non-detected galaxies, binned by other properties (like stellar
mass). \citet{brown15} stack ALFALFA spectra in a sample of
$\sim$25,000 galaxies to study \hi \,
scaling relations fully across the range of gas-rich to gas-poor
galaxies. Still, the stacking studies are limited to making
statistical conclusions about the average properties of galaxies in
each bin.

To improve on the environmental coverage and depth of \hi \, surveys, the
\textit{GALEX} Arecibo SDSS Survey \citep[GASS,][]{catinella10} 
observed a sample of $\sim$800 galaxies with Arecibo until they were
detected in \hi \, or reached an upper limit of $0.015$-$0.05$ in \hi \, gas
fraction (M$_\textrm{\hi}$/\Mst). This sample was the first to
{simultaneously} cover a
substantial volume and measure \hi \, in galaxies across the gas-rich and
gas-poor regimes. One of GASS's main environmental findings
was that massive galaxies (\Mst/\msun>$10^{10}$) in large
halos ($10^{13}$$<$M$_\textrm{h}$/\msun$<$$10^{14}$) have at least 0.4~dex lower \hi \, gas
fractions than those with similar \Mst \, in smaller halos \citep{catinella13}.

In this work, we use the extended GASS sample
\citep[xGASS][]{catinella17}, which includes additional
galaxies at lower stellar masses. Our \hi \, observations are
exceptionally deep and represent the largest sample of galaxies which
probes the gas-poor regime across field and group 
environments. These \hi \, measurements allow us to 
witness the full range of environmental effects on a galaxy's gas,
from the delicate effects of pre-processing in loose groups, to the
conspicuous transformative effects in large clusters. 

In particular, we focus on the 
effects of environment on the gas and star formation properties of
``central'' galaxies. Central galaxies are the {dominant (most
  massive) member in} their group or cluster, {but are sometimes
  also defined as the Brightest Cluster Galaxy (BCG) or Brightest
  Group Galaxy (BGG) (as discussed further in Section~\ref{sec:env})}.
{Central galaxies} usually reside at the center of the group's
dark matter halo {but} can also be found in isolation. Central
galaxies in 
groups grow primarily by mergers and interactions, 
while isolated galaxies experience mostly secular evolution
\citep[e.g.,][and references therein]{lacerna14}.

Central and satellite galaxies are thought to follow different
evolutionary pathways {as they are affected by different
  mechanisms. Satellite galaxies can experience a wide range of
  environmental effects (e.g., ram pressure stripping, tidal
  interactions, etc.) while the evolution of central galaxies is
 more closely tied to their halo mass, involving
fewer mechanisms}, and central galaxies make a
greater contribution to the growth of stellar mass in galaxies
\citep{rp11}. The environmental effects on the \hi \, content of
satellite galaxies are discussed in \citet{brown16} and will
not be {considered further in this work.}

In this work we compare central galaxies in groups and in isolation in
order to identify possible environmental effects on their gas and
star-formation properties. We also consider the effects that group
size (i.e., total dark matter halo mass or multiplicity) and
{local} environmental density {(i.e., the density of
  nearby galaxies within 1 Mpc)} may have on the properties of
central galaxies in our sample. These environmental metrics are some
of the most commonly used when studying the role of environment on
galaxy evolution \citep{blantonberlind07}. Finally, we {make
comparisons between galaxies in different environments} at fixed
stellar mass, since many 
galaxy properties (e.g., star formation, size, luminosity) scale
primarily with stellar mass \citep{kauffmann03}. 

This paper is organised as follows. Section~\ref{sec:sample} describes
and characterizes the sample of galaxies used in this
work. Section~\ref{sec:sfr} and Section~\ref{sec:env} describe our
determinations of star formation rates (SFRs) and environment metrics,
respectively. Section~\ref{sec:results} describes our main results,
and  Section~\ref{sec:discussion} discusses these results and their
implications. We summarize our main conclusions in
Section~\ref{sec:summary}. Throughout this work we use a $\Lambda$CDM
cosmology with H$_0$=70km s$^{-1}$ Mpc$^{-1}$ and $\Omega_\textrm{M}$=0.3.

\section{\texorpdfstring{\MakeLowercase{x}}xGASS Sample}
\label{sec:sample}

{The xGASS survey is an extension of GASS \citep{catinella10} to
  include lower stellar mass galaxies (the GASS-low sample).
}

{The original GASS  sample
  \citep[of][]{catinella13} was selected to have a flat
distribution of stellar mass between
$10^{10}$$\le$\Mst/\msun$\le$$10^{11.5}$ and redshifts
$0.025$$\le$$z$$\le$$0.05$. Each member of the GASS sample was
observed in \hi \, until detected or until an upper limit on the gas
fraction (M$_\text{\hi}$/\Mst) of $0.01-0.05$ was reached.
Since GASS did not target galaxies already detected by ALFALFA, the
observed sample lacked the most gas-rich objects, which needed to be
added back in proportions related to the ALFALFA detection fractions
in the GASS parent sample \citep[see][for complete details]{catinella10}.
This yielded the GASS \textit{representative sample} (760 galaxies),
which was based on statistics estimated from the 40\%
data release of ALFALFA \citep{haynes11} and also included the \hi \,
digital archive \citep{springob05}. With the recent 70\% data
release (AA70}\footnote{obtained from
  \url{http://egg.astro.cornell.edu/alfalfa/data/}})
{of the ALFALFA blind \hi \, survey, 
we revisited the GASS representative sample to just include
homogeneous AA70 observations and updated detection
fractions. It is important to remind the reader that, by construction,
the representative sample still has as flat a stellar mass distribution
as the original GASS sample. The updated GASS representative sample
includes 781 galaxies.}

{Galaxies in the GASS-low sample are selected from a parent
  sample extracted from SDSS DR7 \citep{dr7} having stellar masses
  $10^9$$\le$\Mst/\msun$<$$10^{10.2}$ and redshifts  
between $0.01$$\le$$z$$\le$$0.02$. 208 galaxies selected
randomly were observed with the Arecibo radio telescope. We followed
the same gas fraction limited strategy as GASS, but without imposing a
flat stellar mass distribution.
%
%Following the
%GASS observing strategy, we observed galaxies until detected, or
%until we reached an upper limit of ~0.1-0.02 in gas fraction
%\textbf{[[CHECK NUMBERS]]}. Unlike GASS, no additional constraints on
%the stellar mass 
%distribution of the final sample are included.
This is because at these masses the stellar mass function is flatter
and we sample almost 
equally all the stellar mass range of interest by construction. As in
the case of GASS, for GASS-low we did not re-observe galaxies already
detected by ALFALFA and we created a representative sample following
an analogous procedure. The final xGASS representative sample, which
includes both GASS and GASS-low samples, contains $\sim$1200 galaxies.
}

{No 
environmental or other criteria are imposed on the GASS or GASS-low
sample selections. Complete details of the xGASS
sample selection and its properties are included in
\citet{catinella17}.}

With its large ($3.5'$) beam, the Arecibo \hi \, observations are
susceptible to source confusion if multiple galaxies are nearby each
other on the sky and have similar recession velocities. Each of the
{\hi-detected} xGASS targets are carefully
checked and flagged 
if they have significant confusion from sources within $\sim$2$'$ in
projection (where the beam power drops to half its peak) and within
$\sim$200~\kms \, in recession velocity. We also flag targets with more
distant contaminants if the nearby sources are particularly gas-rich
galaxies. Non-detections in xGASS are not checked for confusion. In
all, we identify {$\sim$10\% of} xGASS targets as significantly
impacted by 
confusion in \hi \, \citep[for complete details see][]{catinella17}.  
%(none of the 17 AA70gcent targets are confused).
In this analysis we only consider the non-confused sample;
Appendix~\ref{sec:others} shows the small changes to our results if
these confused galaxies are not removed.

As will be discussed in Section~\ref{sec:env}, {the xGASS sample
  only contains N=38 non-confused low mass (\Mst/\msun$<$$10^{10.2}$)
  central galaxies in groups.} To improve these
statistics, we searched for additional group central galaxies within
the xGASS mass and redshift range in {the \citet{yang07} group
  catalog (see Section~\ref{sec:env}). We matched these galaxies to
  \hi \, observations from AA70, several of which were already
  included in our xGASS representative sample. However,  we found
  an additional 20 % really 21
  low mass group central galaxies which were not
  included in xGASS, of which 17 are
  detected in \hi \, by AA70, and 3 %really 4, but 1 shredded
  are non-detections. Because this sample of central galaxies is
  nearly complete in \hi, we decided to }
  include these 17 detected sources 
in our analysis and refer to them as the ``AA70gcent'' population. The
potential effects of the {three %really four
un-detected galaxies are small.
% One of
%the four suffers from shredding in the group catalog and is a ``false
%pair'' (see Section~\ref{sec:env}). If the remaining three galaxies
If these}
were observed to have extremely low \hi \, masses, our primary results
would only weakly be affected, as our sample includes 55 low mass
group central galaxies.

{In this work we combine the xGASS and AA70gcent samples, removing
\hi-confused galaxies, those with no estimates of SFR (see
Section~\ref{sec:sfr}), and those not matched in the group catalog
(see Section~\ref{sec:env}). This leaves a final sample of N=1080
galaxies,
of which %759 centrals
there are 234 central galaxies
in groups and 525 in isolation.}

We also use CO(1-0) observations of a subset of the xGASS sample to
estimate their molecular hydrogen (\H2) content. These observations
come from the CO Legacy Data base for the GASS survey
\citep[COLD~GASS,][]{saintonge11} and its low mass {extension
  \citep[COLD~GASS-low,][]{saintonge17}. Analogously to COLD~GASS, the
  low mass extension is a follow-up of a random subset of GASS-low,
  hence its \Mst \, and $z$ intervals are identical for xGASS and
  xCOLD~GASS.}
The xCOLD~GASS sample provides \H2
estimates for $\sim$400 %444
of the galaxies in xGASS.
%(and for none of the AA70gcent sample).
Full details about xCOLD~GASS and its properties are included in
\citet{saintonge17}.

\section{Star formation rate determination}
\label{sec:sfr}

In addition to the observations of the atomic and molecular gas for
galaxies in our sample, we are also interested in quantifying the star
formation 
processes underway in these objects. In an ideal dust-free galaxy, its
ultra-violet (UV) luminosity would be an excellent tracer of recent
($<$100~Myr) star formation. However, 
dust absorbs up to $\sim$$70\%$ of the UV flux and re-emits it at
mid-infrared (MIR) wavelengths, requiring a correction to UV SFRs
\citep{buat99, burgarella13}. Dust 
{emission and absorption vary as} a function of galaxy
properties, so multi-wavelength observations and corrections are
required to determine the total SFRs in a sample of
galaxies (e.g., \citealt{boquien16}).

Toward that end, we generate total SFRs for all galaxies in our sample
using both UV and MIR observations. While there are a variety of
well-tested and statistically robust existing multi-wavelength
star-formation indicators \citep[e.g., the recent UV+MIR SFRs
  from][]{salim16}, the galaxies in our sample are too nearby and too 
{extended to fully} rely on automated {MIR} catalog
photometry, {which is typically best suited for measuring fluxes
  of point-sources.} Our total SFRs are determined
using standard SFR indicators from UV \citep{schiminovich07} and MIR
\citep{jarrett13} luminosity conversions, and include a correction for
stellar MIR contamination \citep{ciesla14}. {All luminosities
  are computed using luminosity distances determined from the SDSS
  redshifts for each source.}

Our UV fluxes come from the \textit{Galaxy Evolution Explorer}
(\textit{GALEX}, 
\citealt{martin05, morrissey07}) which collected UV images and
spectroscopy from 2003 to 2012. We find matches to our sources from
catalogs available in the \textit{GALEX} CasJobs 
interface\footnote{\mbox{\url{https://galex.stsci.edu/casjobs/}}}, 
including both the \citet{bianchi14} Catalog
(BSCAT\footnote{\mbox{\url{https://archive.stsci.edu/prepds/bcscat/}}}),
the \textit{GALEX} Unique Source Catalog
(GCAT\footnote{\mbox{\url{https://archive.stsci.edu/prepds/gcat/}}}),
and the GR6+7 data
release\footnote{\mbox{\url{http://galex.stsci.edu/GR6/}}} 
to obtain observations from the Medium Imaging Survey (MIS, 1500s
exposures) and All Sky Imaging Survey (AIS, 100s exposures). Given
multiple NUV observations of the same 
target, we choose the GCAT measurements over the BSCAT measurements,
and the MIS observations over the AIS observations. {We use the
  ``auto'' flux measurements within Kron-like elliptical
  apertures which are suitable for extended objects.} GCAT-MIS provides 
fluxes for $\sim$$60$\% of our sample, GCAT-AIS provides $\sim$$30$\%,
BSCAT-MIS and BSCAT-AIS together provide $\sim$1\%, GR6+7 provides
$\sim$2\%, and 14 objects do not have any UV flux measurements from
\textit{GALEX}. %Complete details of the UV data are included
%in the xGASS data release \citep{catinella17}.

These \textit{GALEX} catalogs also provide flags on each photometric
measurement, to indicate whether the photometry may be contaminated by
neighbors or if the object has been deblended from a
neighbor. Approximately $80\%$ of our sources have unflagged UV and
are reliable. Even when including the flagged sources, we find good
agreement between these UV fluxes and those measured by \citet{wang11}
for the galaxies in common with this sample. We convert the NUV fluxes
into SFRs using the observed redshifts of the sources, correcting for
Galactic extinction \citep{sf11}, and using the SFR calibration from
\citet{schiminovich07}, as shown in Equation~\ref{eq:sfrnuv}. 

\begin{equation}
\label{eq:sfrnuv}
\textrm{SFR}_\textrm{NUV}[\textrm{\msun \, yr$^{-1}$}]= 10^{-28.165} L_\textrm{NUV}[\textrm{erg/s}]
\end{equation}

\vspace{0.1cm}

Our MIR fluxes come from the \textit{Wide-field Infrared Survey
  Explorer} (\textit{WISE}, \citealt{wright10}), which mapped the
whole sky between 
wavelengths of 3.4 and 22 $\mu$m. Its large angular resolution
($6''/12''$) means that most of its detections are unresolved, and
the AllWISE data
release\footnote{\url{http://wise2.ipac.caltech.edu/docs/release/allwise/}}
includes only profile-fit flux measurements. {While} the
AllWISE stacking {process further blurs the images} (to
  $10''$ and $17''$), most of our targets are still resolved at this
  scale, {so} 
we are unable to use the 
profile-fit measurements. Instead, we perform aperture-photometry on
the atlas images using \textsc{SExtractor} \citep{bertin96},
{and use ``AUTO fluxes measured in Kron-like elliptical
  apertures}. We use the  \wthree \, ($12\mu$m) and \wfour \,
($22\mu$m) images and find that $\sim$90\% of our sources are detected
in \wthree \, and $\sim$60\% are detected in \wfour, which has coarser
resolution and is less sensitive.

To ensure that our MIR flux measurements are not contaminated by
neighbors, we flag all sources which \textsc{SExtractor} identifies as blended,
and also those which have apertures overlapping by
more than $25\%$ with a neighbor that has at least $25\%$ as much flux
as the target (using a geometric algorithm from \citealt{eeover}). This
identifies 46 \wthree \, sources and 22 \wfour \, sources as possibly
contaminated. %Complete details of the MIR photometry are included in
%the xGASS data release \citep{catinella17}.

{We apply the standard
aperture corrections \citep{jarrett13} to our SExtractor ``AUTO''
magnitudes of $\pm$$\sim$$0.03$~mag, and corrections for Galactic
extinction  in \wone \, and \wtwo \, \citep[$\sim$$0.01$~mag,][]{sf11},
but not in \wthree \, and \wfour \, as they are
negligible. We also include a color correction to \wfour \, of
$\sim$$0.1$~mag when \wtwo-\wthree$\ge$$1.3$~mag, as recommended by
\citet{jarrett13}.
}

{
The SDSS redshifts are used to calculate luminosities in each of the
\textit{WISE} bands. 
We also apply a small correction for stellar MIR contamination based on
\wone \, luminosity \citep[][calculated in an analogous
  way to \wthree \, and \wfour]{ciesla14}, and the SFR estimates in \wthree \, and
\wfour \, come from the calibration in \citet{jarrett13}, as shown in
Equations~\ref{eq:sfrw3} and \ref{eq:sfrw4}.
}

\begin{equation}
\label{eq:sfrw3}
\textrm{SFR}_\textrm{w3}[\textrm{\msun \, yr$^{-1}$}]= 4.91\times10^{-10} \times  \left( L_\textrm{w3} -  0.201 L_\textrm{w1} \right)[\textrm{\lsun}]
\end{equation}

\begin{equation}
\label{eq:sfrw4}
\textrm{SFR}_\textrm{w4}[\textrm{\msun \, yr$^{-1}$}]= 7.50\times10^{-10} \times \left( L_\textrm{w4} -  0.044 L_\textrm{w1} \right)[\textrm{\lsun}]
\end{equation}

\vspace{0.2cm}

{For all of the galaxies in our sample with \wfour \,
  detections, the stellar MIR correction was never larger
  than the \wfour \, SFR. For $\sim$50\% of the $\sim$250 galaxies
  detected in \wthree \, and \textit{not} \wfour, the stellar correction
  was larger than the \wthree \, SFR, and so the MIR contribution to
  the total SFR was set to zero. These $\sim$125 galaxies are among the reddest in
  the sample (NUV-r>4.5) and are distributed uniformly across the
  sample volume (with \wthree \, flux errors $\le$3\%). The \wthree \,
  emission in objects like these can be entirely attributed to old
  stellar populations, and not to recent star formation. }

We
verified at this point that there were no systematic differences
between SFR$_\wthree$ and SFR$_\wfour$ estimates for the objects which
were detected and unflagged in both bands. SFR$_\wfour$ is a more
reliable tracer of the SFR; the $12\mu$m luminosity is more
affected by emission from polycyclic aromatic hydrocarbons and old
stellar populations \citep{calzetti07, englebracht08}, and its stellar
MIR contamination correction factor is correspondingly larger.

\begin{figure*}
	\includegraphics[height=4.7cm,valign=t]{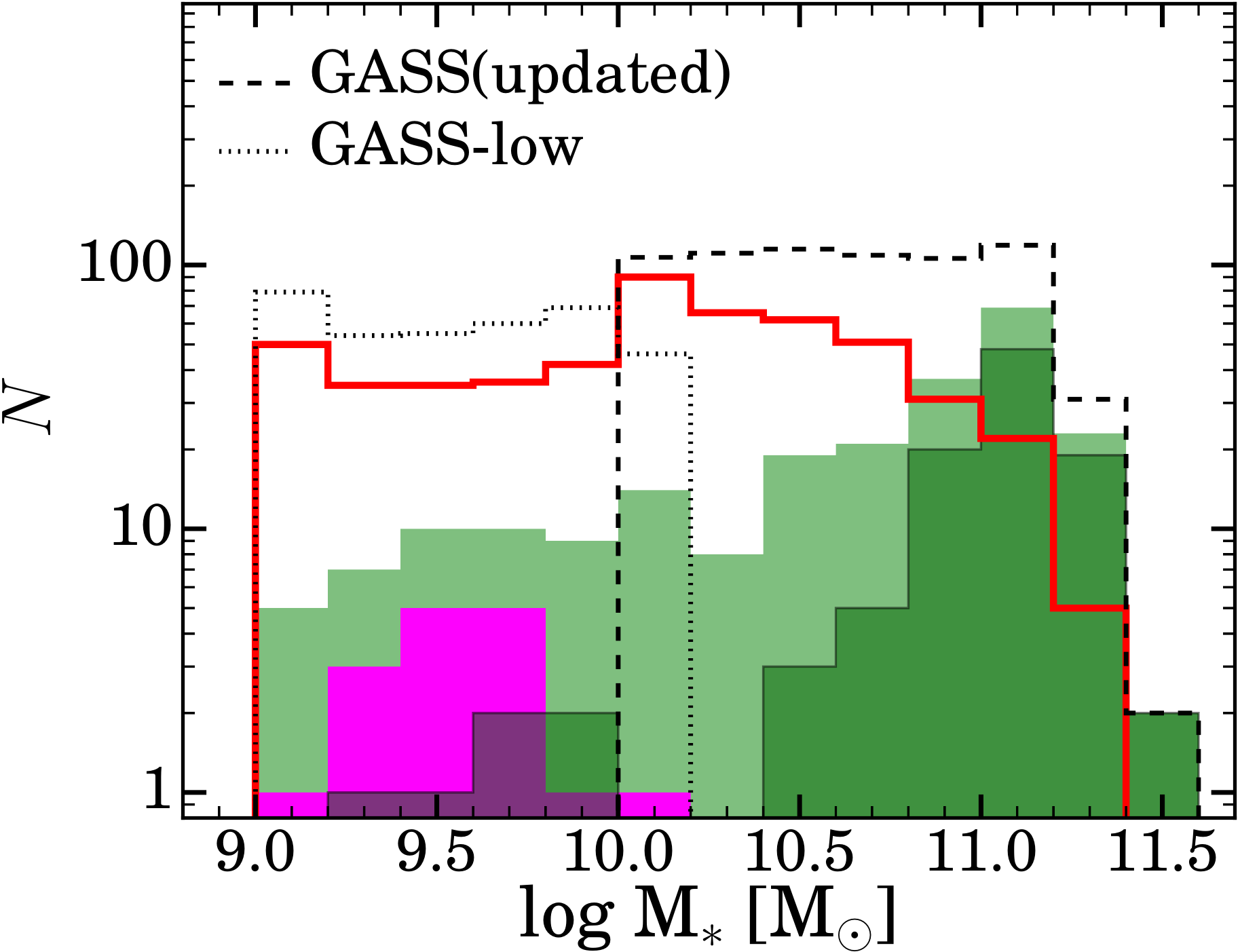}
	\includegraphics[height=4.7cm,valign=t]{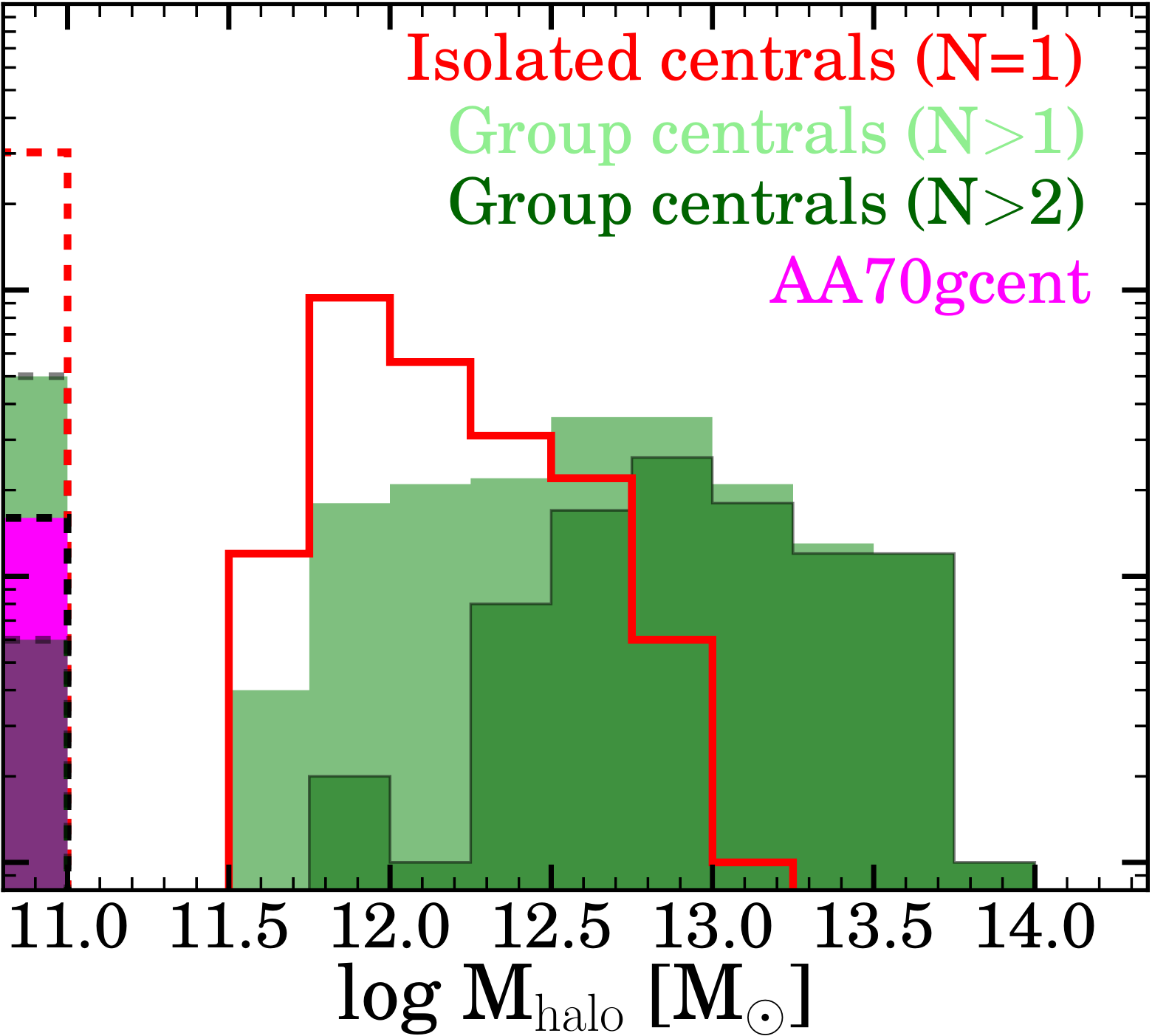}
	\includegraphics[height=4.75cm,valign=t]{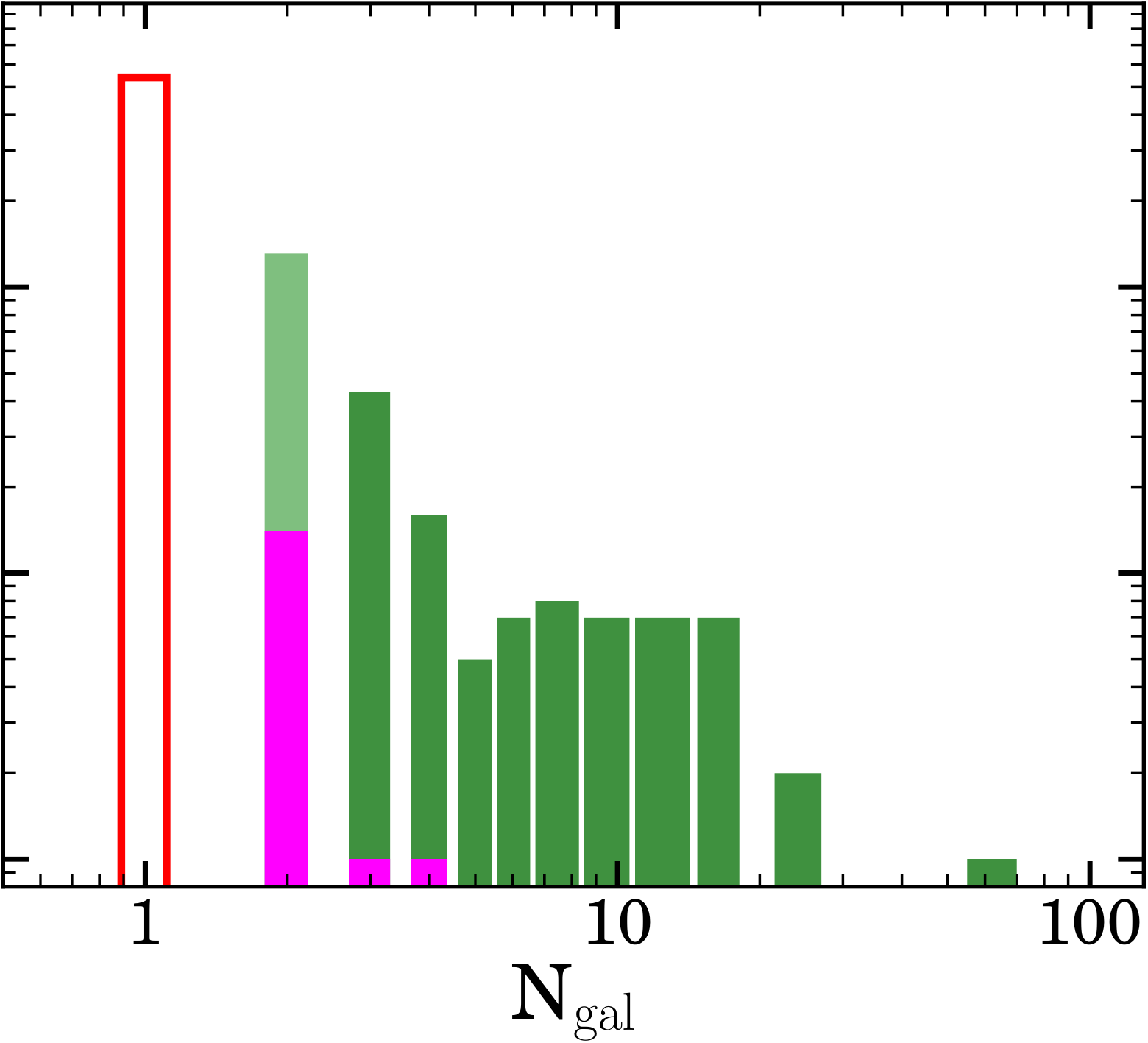}
        %sorry, assymetric??
    \caption{
      Left panel: {histograms of stellar mass for the GASS
        (dashed) and GASS-low (dotted) representative samples are
            shown in black. 
        Central galaxies in our combined sample are shown in groups
        (shaded green) and isolation (red lines).} Note that central galaxies
      are more frequent at larger
      stellar masses. Also shown are the 17 group central
      {galaxies} from AA70 (``AA70gcent''). 
    Center panel: halo mass histogram for central galaxies by
      environment (central galaxies without assigned halo masses are
      shown as dashed histograms at M$_\text{halo}$=$10^{11}$\msun). 
    Right panel: histograms of {group multiplicity (the number
      of group members, N$_\text{gal}$)}.% for
}
    \label{fig:mhist}
\end{figure*}

For galaxies with unflagged MIR and NUV observations, we generate total
SFRs by summing SFR$_\wfour$ (or
SFR$_\wthree$ if necessary) and SFR$_\textrm{NUV}$, as shown in
Equation~\ref{eq:sfrtot}. 

\begin{equation}
\label{eq:sfrtot}
\textrm{SFR}_\textrm{NUV+MIR} = \textrm{SFR}_\textrm{w4} + \textrm{SFR}_\textrm{NUV}
\end{equation}

\vspace{0.1cm}

\noindent
Combined,  
this gives total SFRs for $\sim$70\% of the xGASS sample. For the
remaining $\sim$30\% of sources
where good MIR and NUV observations are not both available, we use SFRs
determined from the SED fits of \citet{wang11}, when available. 
{
For 7 central galaxies in our sample, neither accurate NUV flux
measurements nor SED-fitting SFRs are available. For the three
galaxies with MIR-only detections, we compute MIR-only SFRs, which are
larger than the SFRs from MPA/JHU by
0.1-0.4~dex. We exclude 4 galaxies from our analysis for
which none of the above methods can be applied, mainly as a result of
blended 
sources. These four central galaxies are shown in
Appendix~\ref{sec:others} and their exclusion does not change our
results.
}

\iffalse %shortened/dropped
However, there are 15 objects which lack complete MIR+NUV
  coverage and which do not have good SED fits. Seven of these have
  either NUV or MIR flux measurements alone,} which are used to
estimate their SFR. {Most suffer from blending and source
  confusion in the NUV images, and are not detected in \wthree \, or
  \wfour. Three are central galaxies (i.e., will be included
  in this analysis), and have stellar masses
  $\sim$$10^{10}$\msun. We find that their MIR-only SFRs are larger
  than their MPA/JHU SFRs by 0.1-0.4~dex, so adopt the MIR SFRs. }
%
{There remain 8 galaxies which do not have SFR estimates from
  any method, and are not included in our analysis. Briefly, these are
  cases of either MIR non-detection and NUV blending, or blending in
  both MIR and NUV. Four are satellite
  galaxies, three are isolated central galaxies
  (9.4<log\Mst/\msun<10.6), and one is a group central galaxy
  (log\Mst/\msun=11.0). }
\fi %shortened/dropped

We compared the UV+optical SED SFRs and the NUV+MIR SFRs for the
sources in common and found that the SED SFRs are systematically
{1.49} 
times larger than the NUV+MIR SFRs. We have appropriately corrected
the SED SFRs to be consistent with the full NUV+MIR SFRs. 

To further verify the SFRs determined from NUV+MIR photometry, we
applied this same method to the \hi-selected sample of
\citet{vansistine16}, who 
determined SFRs from narrow-band \Ha imaging of $\sim$1400 nearby
galaxies. For the $\sim$400 %423
 galaxies from their sample which have
reliable and unflagged MIR+NUV observations, we find good agreement 
between the \Ha and MIR+NUV SFR estimates (across 3 orders of
magnitude). The best-fit line (in log-space) between these
measurements has a slope of $0.95$, an intercept of $-0.01$, and a
scatter of $\sim$0.2~dex.
%using: GASS/halpha/halpha_UV_WISE.fits
%with this subset: N=423
% nflag4==0&NUV_FLAGS==0&SFR_NUV>0&SFR_w4c>=0&SFRERR<50&SN4>12&SFRERR>0&NUV_MAGERR<.1&w4_magerr_fixed<.2&AA_1!=54&AA_1!=1009
% X : SFR_w4c + SFR_NUV   (all in log space)
% Y : SFR/1.473
%forward fit (in topcat): 
% m=0.847 b=-0.103
%rev fit:
% m=1.053 b= 0.086
%averages: <m> = 0.95,  <b> = -0.009

\section{Environment metrics}
\label{sec:env}

We use multiple metrics to evaluate the environment of the galaxies
in our sample. Different metrics are sensitive to different aspects of
environment, each affecting galaxy evolution in
different ways.

First, we use the group catalogs of \citet{yang07} to identify whether
galaxies are central (most massive) in their
groups\footnote{{note that we use ``group'' to refer to both
    groups and clusters}}, satellite members of
their group, or not in a group. \citet{yang07} used a halo-based
group finder (including enhancements to typical friends-of-friends
algorithms) to identify groups in Sloan Digital Sky Survey (SDSS) Data
Release 4 \citep[DR4,][]{dr4}. An updated version based on Data
Release 7 \citep[DR7,][]{dr7} has been made available
online\footnote{\mbox{\url{http://gax.shao.ac.cn/data/Group.html}}}.

\iffalse %old long version
\citet{yang07} produce three versions of their DR7 group catalog,
including increasingly more objects from decreasingly reliable
sources. Their ``A'' catalog includes only SDSS DR7 spectroscopic
redshifts (N=599,301), ``B'' adds an additional 3269 redshifts from
other redshift surveys,\footnote{
including 2dFGRS \citep{colless01}, IRAS PSCz \citep{saunders00}, RC3
\citep{RC3}, and KIAS-VAGC \citep{choi10}.}
%2-degree Field redshift survey and
%  Korea Institute for Advanced Study Value-Added Galaxy  Catalog
%][]{colless01, choi10}, 
and ``C'' adds 36,759 ``nearest-neighbor''
redshifts. These ``nearest-neighbor'' estimates assign redshifts to
objects without spectra due to fiber collisions and based on their
nearest neighbors on the 
sky; however, \citet{zehavi02} find that in
$\sim$$40\%$ of cases, these assigned redshifts are significantly
incorrect. We adopt the ``B'' catalog and use the
{stellar mass rankings} to determine which galaxies are central
{(most massive)} in their groups. Although catalog ``B'' 
has fewer objects than ``C'', it is less contaminated by inaccurate
redshifts \citep[see also Section~3.2 of][]{skibba11}. Our results are
not strongly dependent on this choice (see Appendix~\ref{sec:others}).
\fi

%new version:
{
\citet{yang07} produce three versions of their DR7 group catalog,
including increasingly more objects from decreasingly reliable
sources. Their ``A'' catalog includes only SDSS DR7 spectroscopic
redshifts, %(N=599,301),
``B'' adds spectroscopic %3269
redshifts from other} surveys,\footnote{
including 2dFGRS \citep{colless01}, IRAS PSCz \citep{saunders00}, RC3
\citep{RC3}, and KIAS-VAGC \citep{choi10}.}
%2-degree Field redshift survey and
%  Korea Institute for Advanced Study Value-Added Galaxy  Catalog
%][]{colless01, choi10}, 
{
and ``C'' adds %36,759
``nearest-neighbor''
redshifts, which are assigned to objects without spectra (due to fiber
collisions) based on the redshifts of their nearest
neighbors. \citet{zehavi02} find that in 
$\sim$$40\%$ of cases, these assigned redshifts are significantly
inaccurate. We adopt the ``B'' catalog as it is less contaminated by
faulty redshifts than ``C'', but more complete than ``A''
\citep[see also Section~3.2 of][]{skibba11}. Our results are 
not strongly dependent on this choice (see Appendix~\ref{sec:others}).
}

As mentioned in Section~\ref{sec:sample}, due to the scarcity of low mass
($10^9$$\le$\Mst/\msun$<$$10^{10.2}$) group central galaxies in
xGASS %(only N=38),
we supplement the sample with additional
galaxies with \hi \, 
observations from AA70. The xGASS {representative} sample
already includes ALFALFA \hi \, observations {in the correct
  proportions}, 
but we here consider the entire AA70 footprint (within
$0.01$$\le$z$\le$$0.02$) and include the additional 17 low mass group
central galaxies (AA70gcent) in our analysis.

All but 24 of the galaxies in our sample {(xGASS+AA70gcent)} are
matched to members of the 
DR7 group catalog ``B'' of \citet{yang07}. These 24 are typically
unmatched because of their proximity to bright stars or survey edges,
and are not included in our analysis. We also correct
``false pairs'' from this catalog, which are cases where a
single galaxy is broken into multiple objects, each separated by less
than their own size \citep{skibba09}.  After visual inspection, we
find that a similar threshold identifies $\sim$$20$ false pairs in the ``B''
catalog, %^(and an additional 9 in the ``C'' catalog), 
and we remove the
smaller objects of each pair (see Appendix~\ref{sec:yangmod} for a
{list of changes to central galaxies in our sample}). Some of
these were identified as 
central galaxies in groups of N=2, so were corrected to become groups
of N=1. Others were satellite galaxies in groups, so their 
group sizes are reduced by one. In one case (GASS~109081) a central
galaxy in a group (of N=5) has been shredded into three galaxies, so
the group size is corrected to N=3.

Having made these corrections, we can now characterize { the
  xGASS+AA70gcent sample (now excluding galaxies which are confused in
  \hi \, or have no SFR estimate)} in
terms of environmental identities. We find that $\sim$30\% are
classified as satellite galaxies in groups {(and not discussed
  further in this study)}, $\sim$50\% are
identified as isolated central galaxies, and $\sim$20\% are identified
as group central galaxies (i.e. the most massive galaxy in
their group), with at least two group
members. Figure~\ref{fig:mhist} 
shows stellar mass, halo mass, and group multiplicity for the xGASS
and AA70gcent samples, after removing all confused galaxies {and
  those without SFR estimates}. {Half of the groups which host
  central galaxies in our sample have multiplicities (total number of
  galaxies in the group) of N=2, and $\sim$80\% are small groups
  with N$\le$4;
\iffalse
Most of
the groups which host central galaxies in our sample 
have multiplicities (total number of galaxies in the group) of
N=2-4; 
\fi
only above \Mst=$10^{11}$\msun \, are central galaxies
found in large groups with N$>$10 members.}

{At low masses ($10^9$<\Mst/\msun<$10^{10.2}$), our group central
  galaxies are found exclusively in groups with 2-4 members: 
  89\% are found in groups of N=2; 9\% in groups of N=3,
  and 2\% in groups of N=4. This distribution of multiplicities is
  similar to that of the full \citet{yang07} DR7 group catalog, which
  includes $\sim$1300 group central galaxies within this mass
  range. %Of those, $\sim$80\% are in groups of N=2, $\sim$18\% are in
  %N=3, 2.0\% are in N=4, and 0.5\% are in N>4.
}

{
At higher masses (\Mst/\msun$\ge$$10^{10.2}$),
  $\sim$75\% live in groups of N=2-4 and the remainder are in larger
  groups up to N=62. When we discuss results for group central
  galaxies, we include all groups regardless of multiplicity. At
  low stellar masses, our group centrals are dominated by N=2 groups,
  while at high stellar masses there are larger groups. We will
  distinguish the N=2 and N>2 populations to show which types of
  groups are driving the trends we see.}

\iffalse
groups of N=2-4 members, in
  order to make consistent comparisons across high and low stellar
  masses.  This multiplicity
  restriction does not mean that our findings apply to all groups of
  N=2-4 members, as our sample is not selected based on group
  multiplicity and only includes six low mass group centrals with
  N=3-4. Rather, our 58 low mass 
  group centrals have a similar distribution of multiplicities as the
  much larger \citet{yang07} DR7 group catalog.
\fi

The group halo masses are assigned 
with an abundance-matching method and are only available for massive
halos with M$_\text{halo}$$\gtrsim$$10^{11.5}$\msun \, \citep{yang07};
smaller halos do not have mass estimates.

It is worth noting that identifying galaxy groups (with $\lesssim$10
members) and assigning ``central'' or 
``satellite'' identities to galaxies is increasingly difficult for
smaller groups \citep{berlind06}. Studies using 
mock catalogs have  shown that { it is especially difficult to
  identify the galaxy at the center of the
  halo (either as most massive or brightest) in the dark matter halos
of small groups}. \citet{skibba11} used mock 
catalogs to show that the {BGGs} in $\sim$$40$\% of
low mass halos ($10^{12}$-$10^{13}$\msun) are not {located at
  the center of the halo (i.e., the galaxy at the center of the
  group's dark matter} halo is not the 
brightest). The fraction of larger halos ($>$$10^{13}$\msun) with this
discrepancy decreases to $\sim$$25$\%. Similarly,
\citet{vonderlinden07} used SDSS observations of 625 galaxy
clusters to show that $\sim$$50$\% of the BCGs are not at the center
of their cluster density fields. {We bear these challenges in
  mind with our simple distinction between central and satellite
  galaxies based on their stellar mass ranking within their
  group.}

Further {complicating this picture is the possibility that}
central galaxies in small groups may experience multiple 
transitions between isolation and group environments. If two small
isolated galaxies interact and become gravitationally bound, one will
become a group central and the other a satellite. If they later merge,
the resulting galaxy will become ``isolated'' again. \citet{park2008}
find that a significant fraction of isolated galaxies are actually the
products of recent mergers, and that recent mergers are even more
common among luminous isolated galaxies. These types of difficulties
are inherent in any attempt to study the smallest galaxy groups, and
must be kept in mind.

In addition to the group membership and environmental identity of the
galaxies in our sample, we have also estimated the local density
in fixed apertures around each object. This calculation is made using
a sample of galaxies from SDSS DR7 \citep{dr7} with
\Mst/\msun$\ge$$10^9$ and which fully 
encompasses the ALFALFA
footprint \citep[see Section~2 of][for more details]{brown16}. The
local density around 
each target is determined by counting the number of galaxies within a
1~Mpc (projected) radius and 1000~\kms \, velocity difference. The
projected densities are 
calculated in units of Mpc$^{-2}$ and include the target galaxy itself
(so have a minimum value of $1/\pi$~Mpc$^{-2}$).

%pre-placed
\begin{figure*}
\centering
\includegraphics[width=0.99\textwidth]{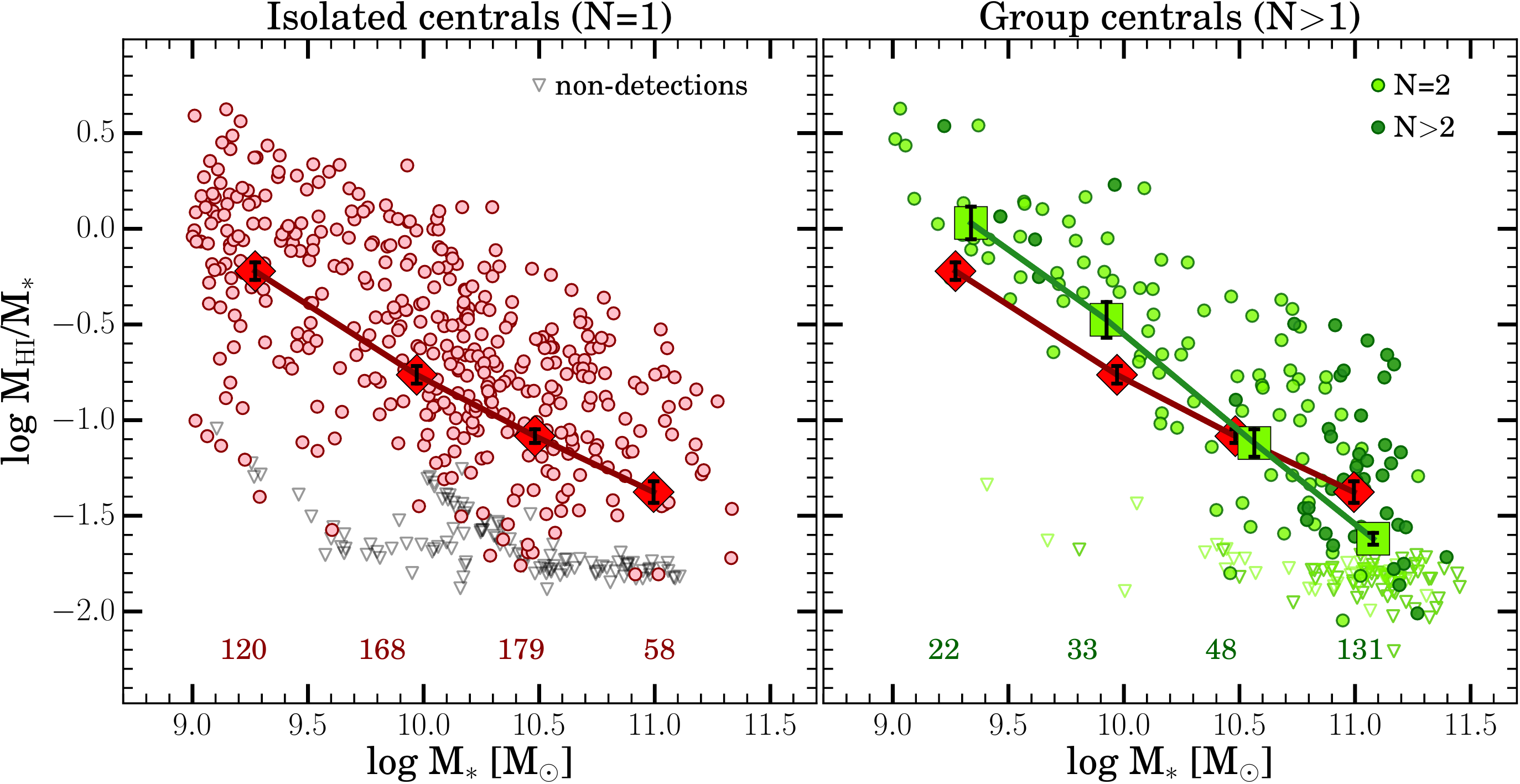}
    \caption{
   On both panels, \hi \, gas fraction of our central galaxies is
   plotted as a function of stellar mass. For this and all
     subsequent figures, average values within bins are shown at the
     average x- and y-values of points within that bin, and error bars
     show standard error {of} the mean. Non-detections are
     included in averages at their upper limits.
   Left panel shows isolated central galaxies while right panel shows
     group central galaxies {(N=2 in dark green and N$>$2 in
       light green);
     \iffalse in groups of N=2-4; 
     \fi the average relations
     for isolated central galaxies are shown as large red diamonds in
     both panels and the averages for group centrals (at all
     multiplicities) are shown as large green squares.}
   Open  triangles show upper limits of non-detections; no
     \hi-confused sources are included.
   Numbers at the bottom of both panels indicate how many galaxies
     were averaged in each bin.
   Heavy coloured lines connect averages of the logarithm of the \hi \, gas
   fraction ($\langle$log~\mhi/\Mst$\rangle$) in bins of stellar mass.
  At low stellar masses (\Mst/\msun$<$$10^{10.2}$), central galaxies in
     groups (green symbols) are rarely gas-poor and have \hi \, gas
     fractions which are on average $\sim$$0.3$~dex larger than
     those in isolation (in red, both panels). 
    }
    \label{fig:envHI}
\end{figure*}

\begin{figure*}
\centering
\includegraphics[width=0.99\textwidth]{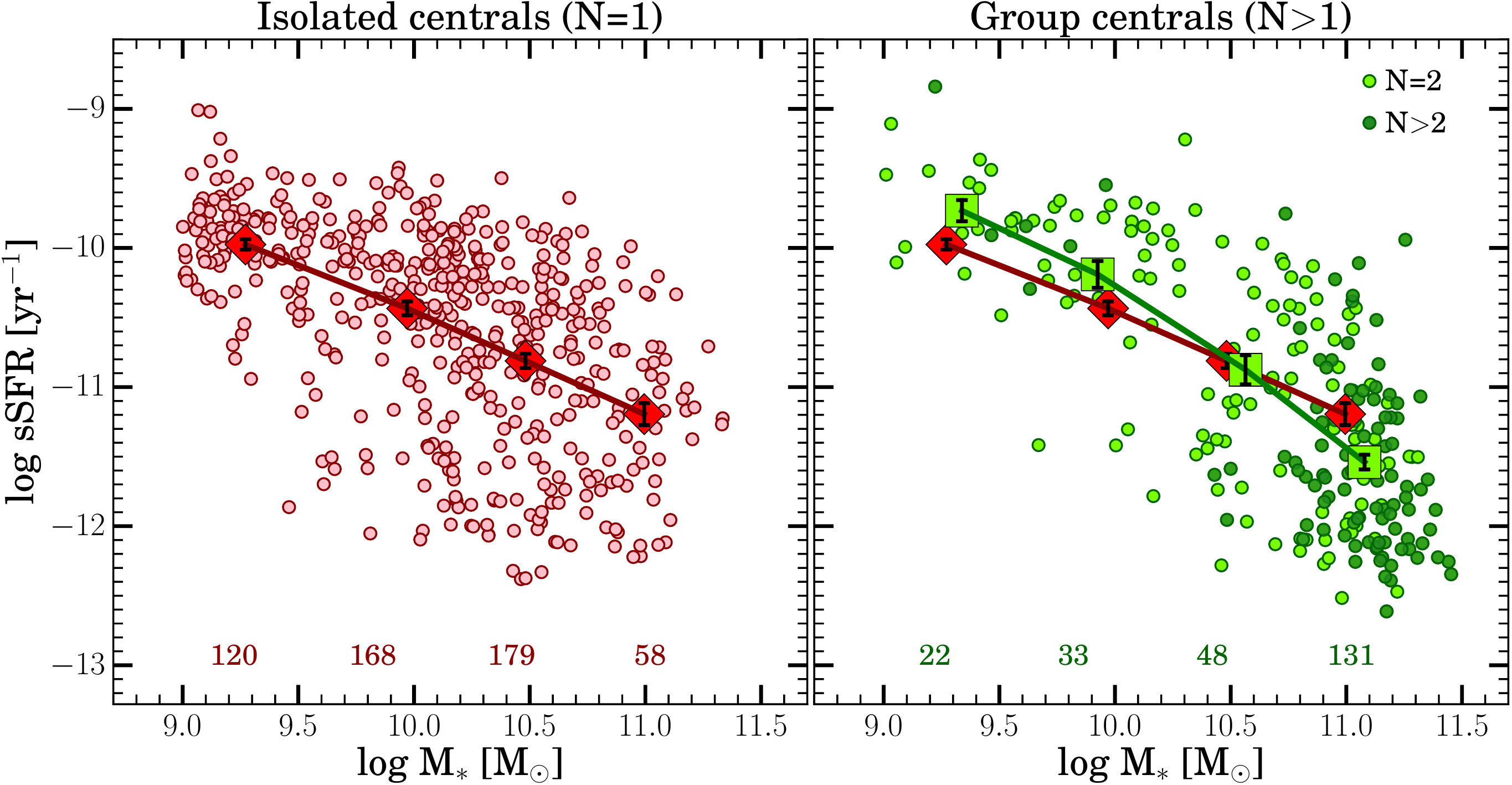}
    \caption{
   Specific star formation rate is plotted against stellar mass for
     central galaxies in both environments. The same galaxies are
     shown with the same colour-coding and averaging as in
     Figure~\ref{fig:envHI}.
   The low mass group central galaxies (shown in green) have sSFRs
   which are elevated by $\sim$$0.2$~dex compared with isolated
   central galaxies (shown in red, with average relation on both
   panels).
    }
    \label{fig:sfr}
\end{figure*}

As a final check, we also verify that we are not being affected by
un-detected satellites around galaxies near the magnitude limit of our
parent sample. Satellite galaxies are typically $\sim$2.5~mag
optically fainter than their central galaxy, so we would be unable to 
detect any satellite galaxies around a central galaxy which is only
$\sim$2.5~mag brighter than our magnitude limit
\citep{lacerna14}. This could lead to an artificial increase in the
fraction of isolated central galaxies in the faintest 2.5~mag of the
sample.

To verify that this effect does not bias our results, we create a
``bright'' sub-set of galaxies which only includes objects $2.5$~mag
brighter than the SDSS {spectroscopic} survey limit. At low masses
(\Mst/\msun$<$$10^{10.2}$), $\sim$55\% of our central galaxies are
included in this ``bright'' subset, as are $\sim$75\% of central
galaxies at high masses (\Mst/\msun$>$$10^{10.2}$). The isolated
central galaxies in this ``bright'' sample are more confidently
isolated galaxies, and are not artificially isolated because their
satellites are too faint to be detected. This ``bright'' sub-set shows
the same main relations and trends as the full sample, and is shown in
detail in Appendix~\ref{sec:others}. 

\begin{figure*}
\centering
\includegraphics[width=8.9cm,valign=t]{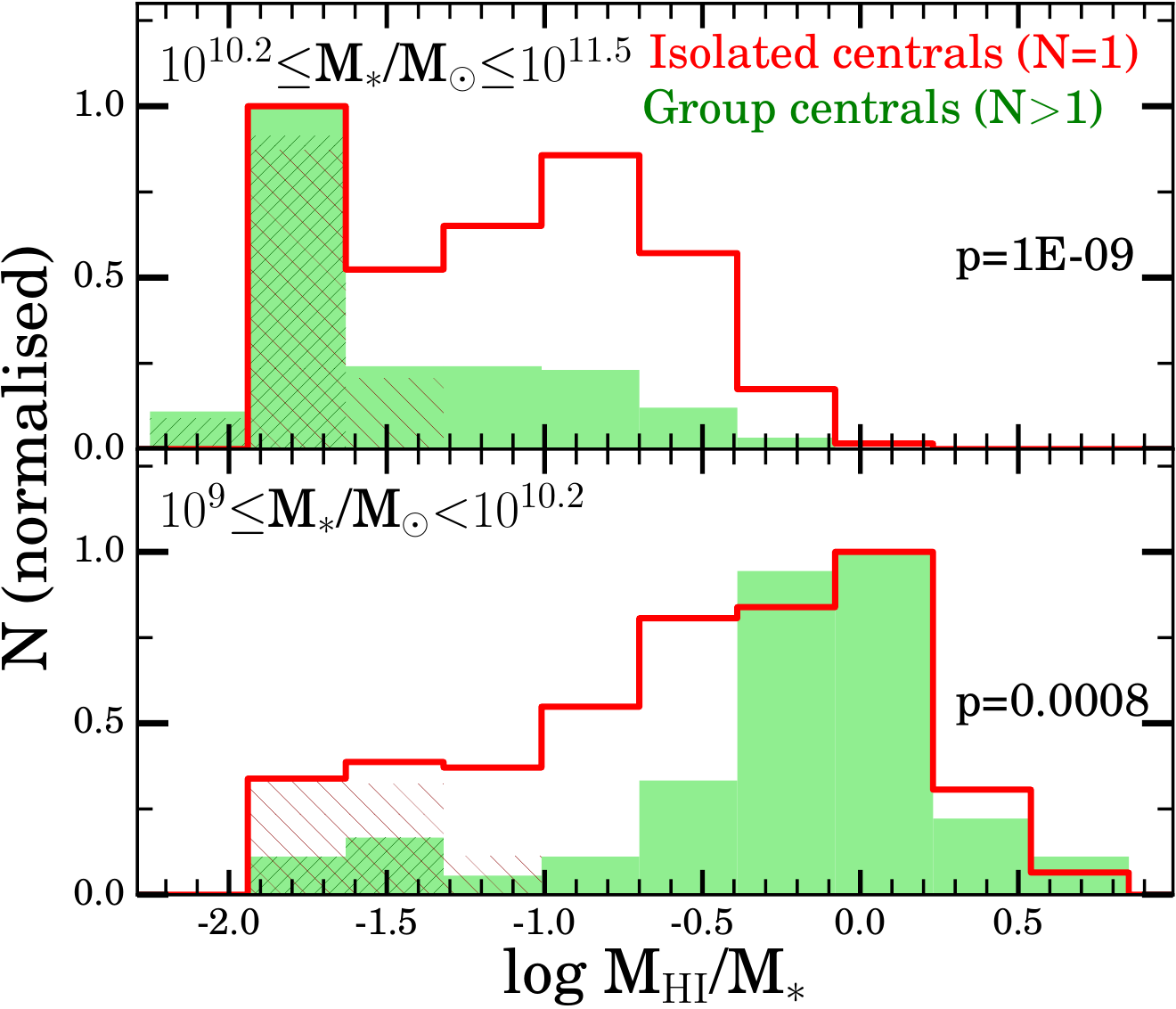}
\includegraphics[width=7.88cm,valign=t]{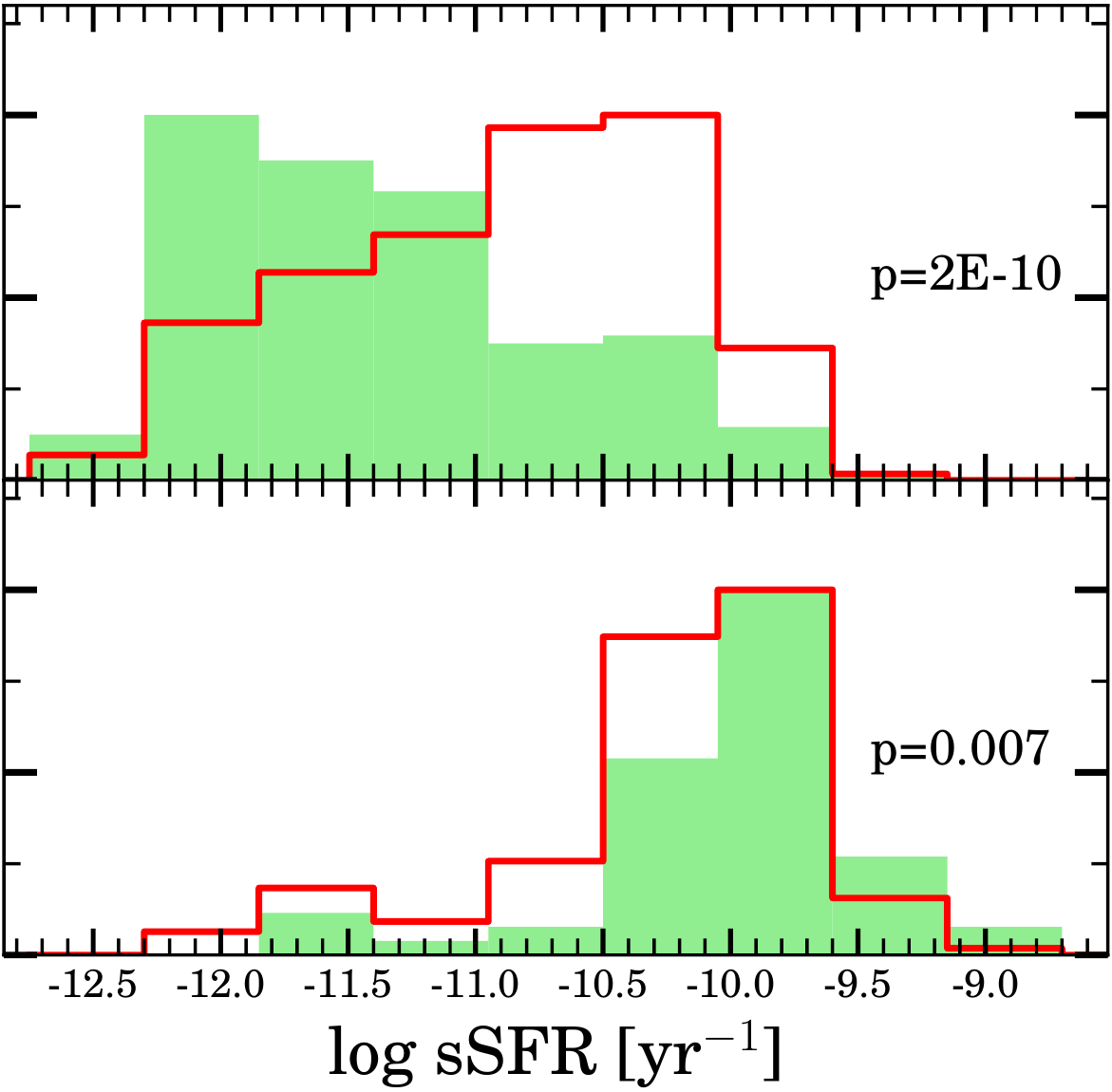}
%smaller:
%\includegraphics[width=7.9cm,valign=t]{dist_hist2_gf-crop.pdf}
%\includegraphics[width=6.99cm,valign=t]{dist_hist2_ssfr-crop.pdf}
    \caption{
   Each panel shows the \hi \, gas fraction or sSFR 
      distributions in large bins of stellar mass (ranges shown at top left).
   Group central galaxies are shaded in light green and 
     isolated central galaxies are heavy red lines. All histograms are
     normalised to have the same peak value.
   In the left column, non-detections in both environments are shown
      as shaded regions; no \hi-confused sources are included in any panel.
   Also shown are the p-values from a two-sided %asymptotic
     Kolmogorov-Smirnov test comparing the group and isolated
     central galaxies (including \hi \, non-detections at their upper
     limits).
   These distributions quantify the sSFR and \hi \, differences
   between central galaxies in groups and in isolation.
    }
    \label{fig:dist}
\end{figure*}

\section{Results}
\label{sec:results}

\iffalse %now included in Sample section
%
The following section describes our primary observational
results. Starting from the total xGASS+AA70gcent sample 
%of {$\sim$1200} galaxies 
(see Section~\ref{sec:sample}), we remove
galaxies which are confused in \hi, which have no estimate of SFR (see
Section~\ref{sec:sfr}), or which are not matched in the group catalog
(see Section~\ref{sec:env}). This reduces our sample
to %N=1076, of which 776 
{$\sim$1100, of which $\sim$800}
are central galaxies (in groups or isolation).
%
\fi

\subsection{Gas rich central galaxies in small groups}
\label{sec:main}

Our main goal is to understand the effects of the group environment
on the \hi \, properties of central galaxies. Toward that end,
Figure~\ref{fig:envHI} shows the \hi \, gas fraction (M$_\text{\hi}$/\Mst)
as a function of 
stellar mass for central galaxies in our sample, 
separated between isolated (left panel) and group (right panel)
environments. Across stellar masses, galaxies in both environments
fully populate the $\sim$1.5~dex of \hi \, gas fraction parameter space,
but there are significant differences between the distributions.
The average values of \hi \, gas fraction in each stellar mass bin
show a general decrease as a function of stellar mass, with lower mass
galaxies being more gas-rich in {both} 
environments, as has been previously found \citep{kannappan09,
  catinella10, cortese11, huang12a, brown15,brown16}.

However, at low stellar mass
($10^{9}$$\le$\Mst/\msun$<$$10^{10.2}$), the central galaxies in
groups (shown as large green {squares}) have $0.3$~dex \textbf{larger}
average \hi \, gas fractions than isolated central galaxies of the
same mass (red diamonds), and are rarely found below the average value
of the isolated galaxies. {At these low masses, $\sim$90\% of the
  groups have multiplicity N=2.} We include non-detected galaxies at
their upper limits averaging the \hi \, gas fractions. At moderate masses
($10^{10.2}$$\le$\Mst/\msun$<$$10^{10.8}$) the group central galaxies
have similar average \hi \, gas fractions to isolated galaxies, and
for \Mst/\msun$\ge$$10^{10.8}$ they are more gas poor than isolated
galaxies.

%only populates groups of N=2-4, so we apply the same
%  range to the high mass group central population. Larger samples of
%  low mass group central galaxies could have groups with N=5 or 6

In addition to the \hi \, relations, we can also test for differences
between the specific SFR (sSFR) of the central galaxies in groups and
in isolation. \hi \, and star formation are closely related
\citep[e.g.,][]{kennicutt98}, and we expect gas-rich 
galaxies to have higher sSFRs. 
\iffalse
Furthermore, sSFR measurements are not
susceptible to confusion when galaxies have nearby companions or are
in a group environment.
\fi

Figure~\ref{fig:sfr} shows the relationship between our sSFR estimates
(described in Section~\ref{sec:sfr}) and stellar mass for our 
sample, divided by environmental identity. The average trends for sSFR
in each environmental type are the same as those seen in the \hi \, gas
fraction plots. When comparing group central galaxies with isolated
galaxies, those with low stellar mass show larger sSFRs by
$0.2$$-$$0.3$~dex.

To better quantify the differences between group and isolated
centrals, Figure~\ref{fig:dist} shows the distributions of \hi \, gas
fraction and sSFR in bins of stellar mass. In the low mass bin, the
isolated central galaxies have a larger gas-poor population than the
group central galaxies. Also shown are the p-values of a two-sided
Kolmogorov-Smirnov test which show that the group and isolated central
distributions are significantly different. At low masses, the
difference in average \hi \, gas fraction between the group and
isolated central galaxy populations is driven by a
\textit{near-absence of gas-poor low mass group central galaxies.}

\subsection{Consistency with \hi \, stacking and sSFR relations in
  larger samples} 
\label{sec:reinforce}

Given the inherent difficulties associated with identifying the
smallest groups of galaxies
(see Section~\ref{sec:env}) and the relatively small number of
galaxies in our sample, %(55 low mass group central galaxies with
%$10^{9}$$\le$\Mst/\msun<$10^{10.2}$), 
we next explore ways to verify the properties of
these low mass group central galaxies with larger statistical
samples.

First, to reach beyond the limits of our sample of \hi-detected
galaxies, we use an \hi \, spectral stacking technique on a much larger
sample of galaxies drawn from the ALFALFA blind \hi \, survey. While the
survey depth is 
insufficient to detect individual galaxies in the gas-poor regime,
stacking many \hi \, spectra can produce a statistical detection below
its nominal sensitivity limit. We compare with the sample of
N$\sim$25,000 \hi 
\, spectra {from and following the methodology} of
\citet{brown15} {again using the}
\citet{yang07} DR7 group B catalog to test whether this same
difference is observed.  We include any central galaxies that match
our sample selection ({i.e., between $10^9$$\le$\Mst/\msun$<$$10^{10.2}$ and
  0.01$\le$z$\le$0.02, or between $10^{10}$$\le$\Mst/\msun$\le$$10^{11.5}$ and
  0.025$\le$z$\le$0.05}) find {$\sim$2400 in groups} %in small groups (of N=2-4) 
and $\sim$11,000 in isolation.

To avoid possible \hi \, confusion in the stacking process, galaxies are
not included in stacks if they have a neighbor within a projected
separation of $2'$ and velocity difference smaller than $\pm200$~\kms
regardless of their optical colour. {This threshold is quite
  conservative, as the Arecibo beam power is at half its peak at
  this radius and red galaxies would be unlikely to contribute any \hi
  \, flux to the observed \hi \, signal. Nonetheless, this }
confusion criterion eliminates $\sim$$1\%$ of isolated central
galaxies and {$\sim$15}\% of galaxies in groups, %of N=2-4, 
but still gives a
statistically robust sample. As an additional test, we used even
more aggressive thresholds of $3'$ and 300~\kms, and the results are
unchanged. While the confusion-cleaned stacks include fewer objects,
the results are more reliable.

\begin{figure}
\centering
\includegraphics[width=0.99\columnwidth]{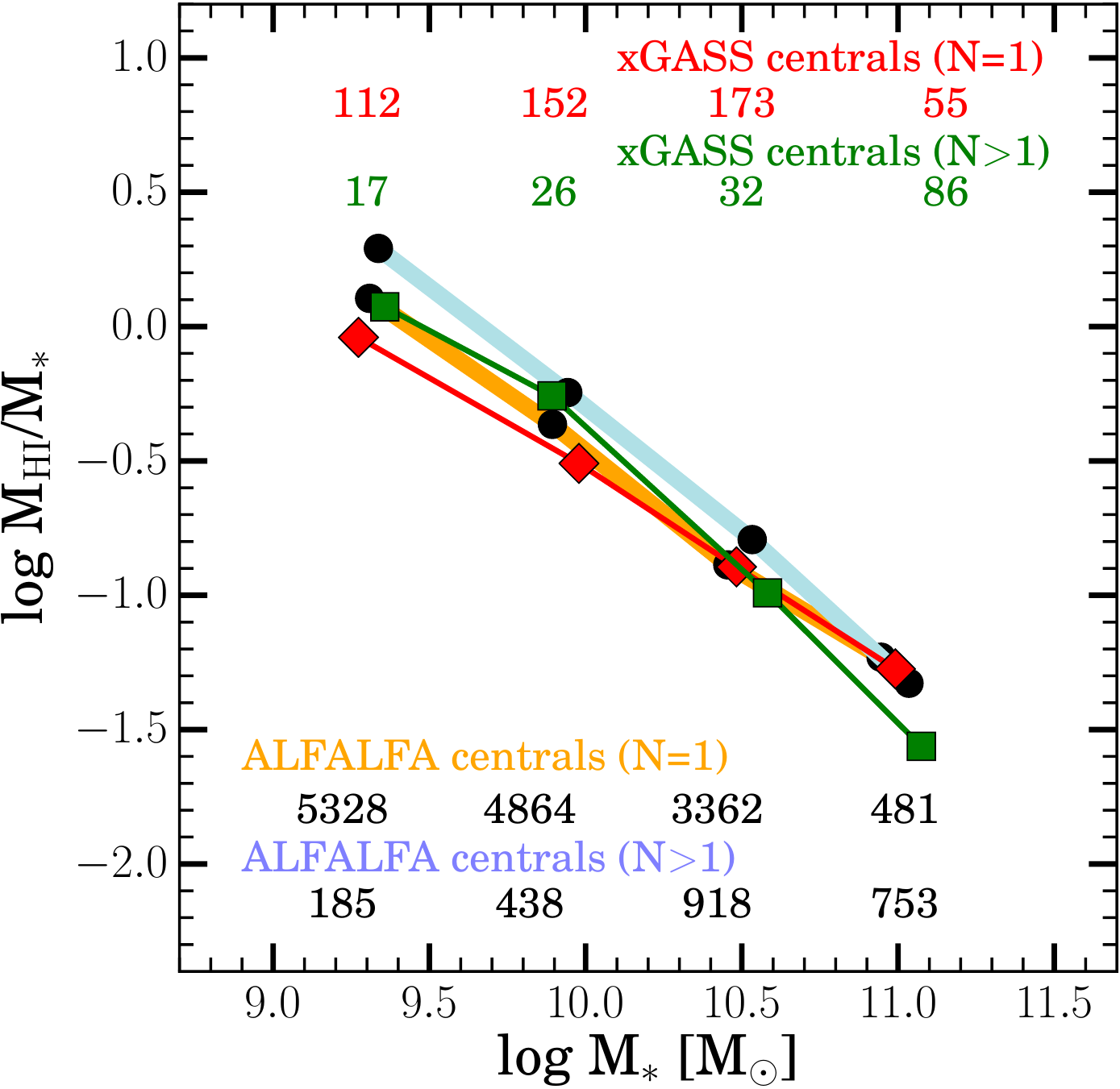}
    \caption{
  The logarithm of the stacked \hi \, gas fraction
  (log$\langle$M$_\textrm{\hi}$/\Mst$\rangle$) is plotted as a
    function of stellar mass for central galaxies in isolated (N=1,
    thick orange line)
    and group ({N$>$2}, light blue line) environments. 
  Isolated (red line and points) and group (green line and points)
    central galaxies from our sample are also shown (confused galaxies
    are removed 
    from both samples). Jack-knife estimates of uncertainties are not
    plotted on individual points, but are comparable to the size of
    the symbols.
  The number of galaxies in each stacked bin is shown at the bottom.
   The relations from the \hi \, stacking sample show the same difference
   between the gas fractions of low mass central galaxies in groups
   and in isolation.
    }
    \label{fig:envHIstack}
\end{figure}

{As described in \citet{brown15}, we start the stacking
  process by shifting individual HI spectra (both detections and
  non-detections) to a common rest-frame frequency. Next we weight
  each galaxy's spectrum by its stellar mass, to stack in units of
  \hi \, gas fraction \citep[see][]{fabello11}. In each stack the
  resulting spectrum is a strong detection, with signal-to-noise
  ratios (calculated as the peak flux divided by the rms noise)
  between 12 and 74.}

Figure~\ref{fig:envHIstack} shows the \hi \, gas fraction as a function of 
stellar mass for central galaxies in isolation and groups, as
measured by stacking the \hi \, spectra of galaxies in each bin. We stack
the xGASS spectra in the same manner as the ALFALFA sample (including
the same $2'$ and 200~\kms \, threshold cuts for confusion, which
reduce the number of xGASS objects in each bin compared with
Figure~\ref{fig:envHI}). Because 
stacking is inherently a linear process, Figure~\ref{fig:envHIstack} 
shows the logarithm of the average \hi \, gas fraction
(log~$\langle$M$_\textrm{\hi}$/\Mst$\rangle$), while our  
previous Figure~\ref{fig:envHI} showed the average of the logarithm of
the \hi \, gas fraction ($\langle$log~M$_\textrm{\hi}$/\Mst$\rangle$).

\begin{figure}
\centering
\includegraphics[width=0.99\columnwidth]{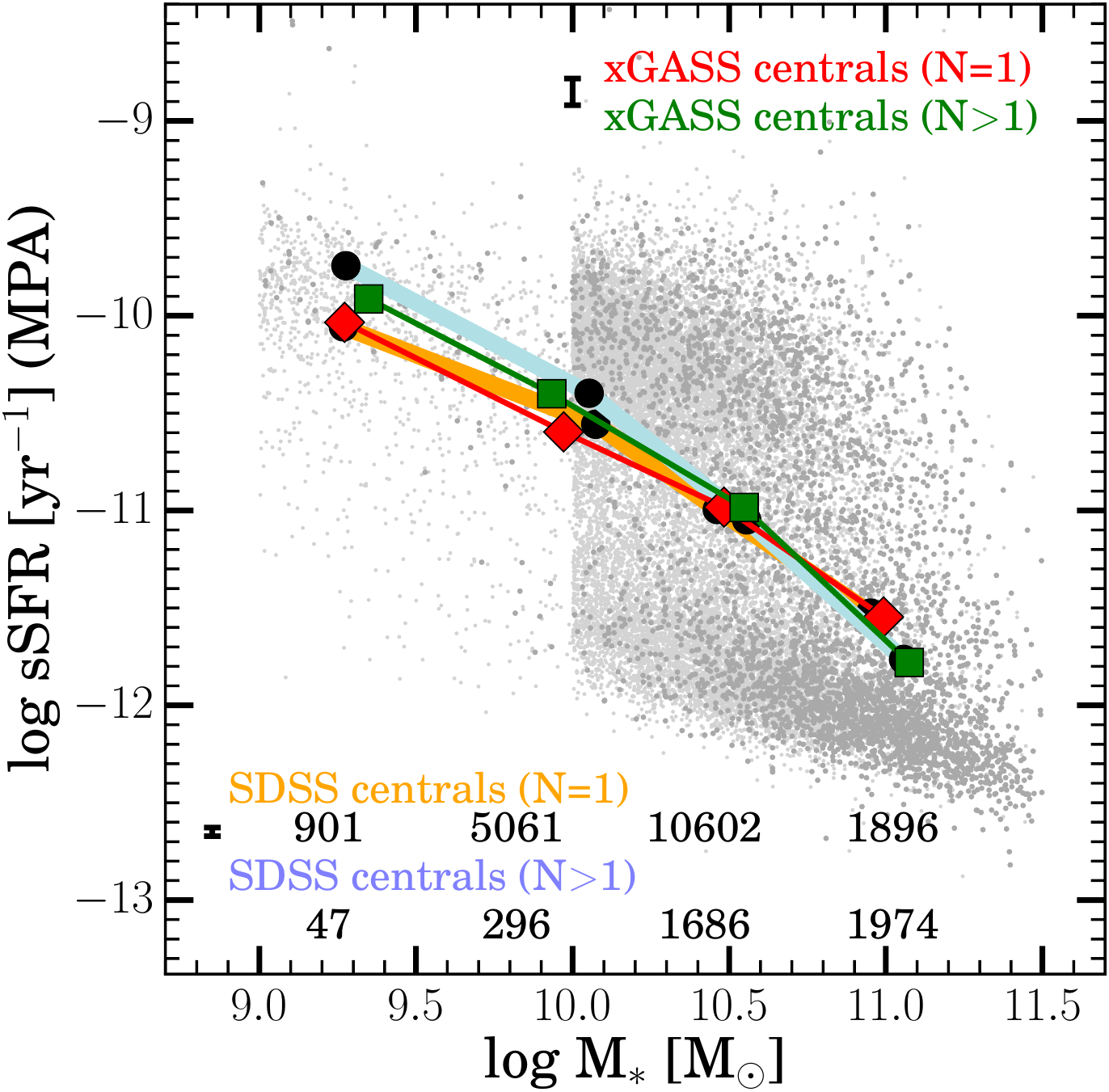}
   %made by SSFRM.py (NOT the version in SSFRM_all.py!!!!)
    \caption{
   Average values of specific star formation rate (from the MPA/JHU
   catalog) are shown in bins of
     stellar mass for central galaxies.
   Relations from our sample are shown
   with thin lines as in Figure~\ref{fig:envHIstack}. The SDSS central
   galaxies are shown as light grey (isolated) and dark grey
   (group) points, their average relations are shown as thick shaded
   lines, and the number of galaxies in each bin are shown at the 
   bottom. Heavy 
   black error bars show the typical values of the standard error {of}
   the mean for each sample.
  The relations from the large comparison sample are in 
    agreement with those from our sample, and again the low
   mass group central galaxies show larger sSFRs compared with
   those in isolation.
    }
    \label{fig:sfr2}
\end{figure}

There is good agreement between the trends seen in the stacked xGASS
\hi \, gas fraction relations and those from the ALFALFA sample of
\citet{brown15}. In both samples, the low mass 
{($10^9$$\le$\Mst/\msun$<$$10^{10.2}$)} group central 
galaxies have \hi \, gas fractions which are $\sim$0.2~dex higher than
isolated central galaxies of similar mass. 
In the highest mass bin {($10^{10.8}$$\le$\Mst/\msun$\le$$10^{11.5}$)}, the
group central galaxies in xGASS have a lower average \hi \,
gas fraction by $\sim${0.25}~dex than those in the stacked sample. This
offset results from the difference in stellar mass
distributions between xGASS (selected to have a flat distribution of
\Mst) and the ALFALFA sample (volume-limited, with a {steeper}
power law decline at these masses). Within this bin, the xGASS
galaxies {have $\sim$0.1~dex lower sSFRs} than those in the
ALFALFA stacking sample. {The disagreement between stacked xGASS
  and ALFALFA group central galaxies at high masses is a result of
  different sample selection, but both samples show an offset
  between group and isolated central galaxies at low masses.}

\iffalse %drop, I think.
As further confirmation we replicate our main results
using the other versions (``A'' and ``C'') of the DR7 group catalog
from \citet{yang07}. We find similar trends in \hi \,
gas fractions for the central galaxies, as shown and discussed further
in Appendix~\ref{sec:others}.
\fi

%preplace
\begin{figure}
\centering
\includegraphics[width=0.99\columnwidth]{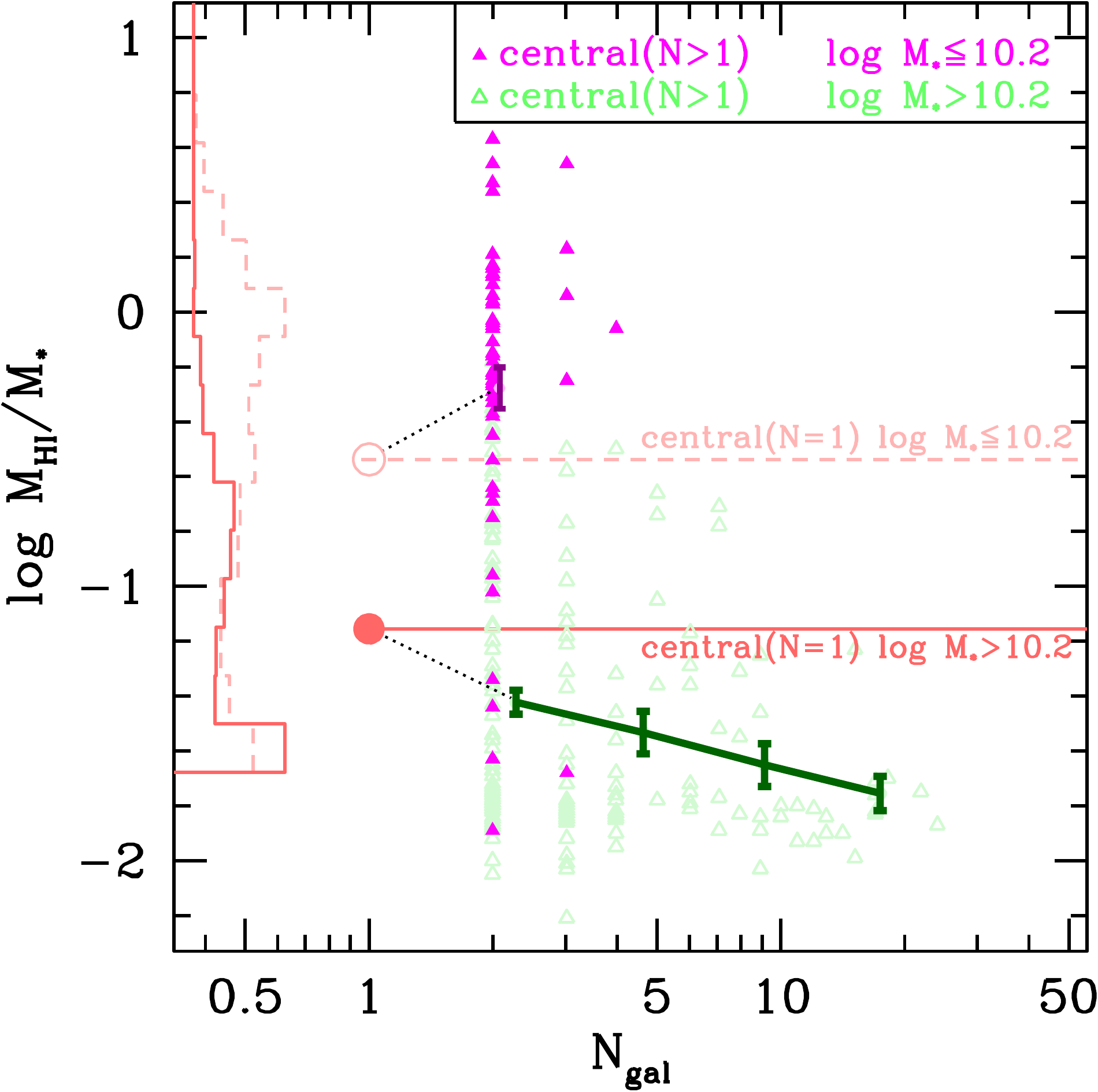}
    \caption{
  \hi \, gas fraction
 as a function of N$_\textrm{gal}$ for central galaxies.
  Isolated central galaxies {(N=1)} %(N$_\textrm{gal}$=1)
   are shown as
     red histograms against the y-axis in dashed (low mass) and solid
     (high mass) lines. Their average \hi \, gas fractions are shown as
     open points and horizontal lines for comparison.
  Group central galaxies {(N>1)} are shown as filled magenta triangles (low
    mass) and open green triangles (high mass), and average
    trends are shown with connected points. A dotted black line
    connects the relations between isolated and group environments.
  %{The grey-shaded area shows the high mass group central galaxies
  %    which have N$_\textrm{gal}$>4, and are not directly comparable
  %    to low mass group centrals in smaller groups.}
  The trend among high mass group centrals
    seems to smoothly continue up to the value of the isolated
    centrals. However, the low mass group centrals are
    more \hi \, rich than comparable isolated galaxies.
    }
    \label{fig:Ngal}
\end{figure}

To confirm our observed sSFR differences in a larger sample of
galaxies, we use the
MPA/JHU galaxy catalog and the DR7 group catalog of
\citet{yang07}, selected within the same stellar mass and redshift
ranges as our sample. In this comparison, we also use the
MPA/JHU sSFR estimates for galaxies in our sample, for
consistency between SFR calibrations. Figure~\ref{fig:sfr2} shows 
the relationship between sSFR and stellar mass, binned in the same way
as Figure~\ref{fig:sfr}. The behavior of central galaxies in isolation
and small groups is well-matched between xGASS and the large MPA/JHU
sample. This large sample of $\sim$32,000~galaxies contains $\sim$3000
group centrals, $\sim$300 %335 
of which populate the lowest two bins of stellar
mass.

To summarise, the low mass group central galaxies appear consistently
elevated by 
$0.2$$-$$0.3$~dex in \hi \, gas fraction and sSFR, whether measured in our
sample, in the \hi \, stacking analysis, or in the MPA/JHU catalog. This
widespread agreement further confirms that these galaxies are
unusually gas-rich and star-forming.

Next we explore whether other properties of the low mass group central
galaxies (or their groups) might help explain their \hi \, and sSFR
properties. We consider the role of the group size (e.g., halo mass
or multiplicity of members), the proximity of and star formation in
their nearest satellite galaxy, and correlations with large-scale
density measurements.

\subsection{Trends with group multiplicity}
\label{sec:halo}

Halo mass is an important property driving group evolution, and is
closely related to hydrodynamical feedback effects like ram pressure
stripping. However, it is very difficult to estimate the total dark
matter halo mass in the small groups of our low mass central
galaxies. In particular, the DR7 group catalog of 
\citet{yang07} does not provide any halo mass estimates for groups
{which have a central 
galaxy below \Mst/\msun<$10^{10}$.}

%pre-placed
\begin{figure}
\centering
\includegraphics[width=0.99\columnwidth]{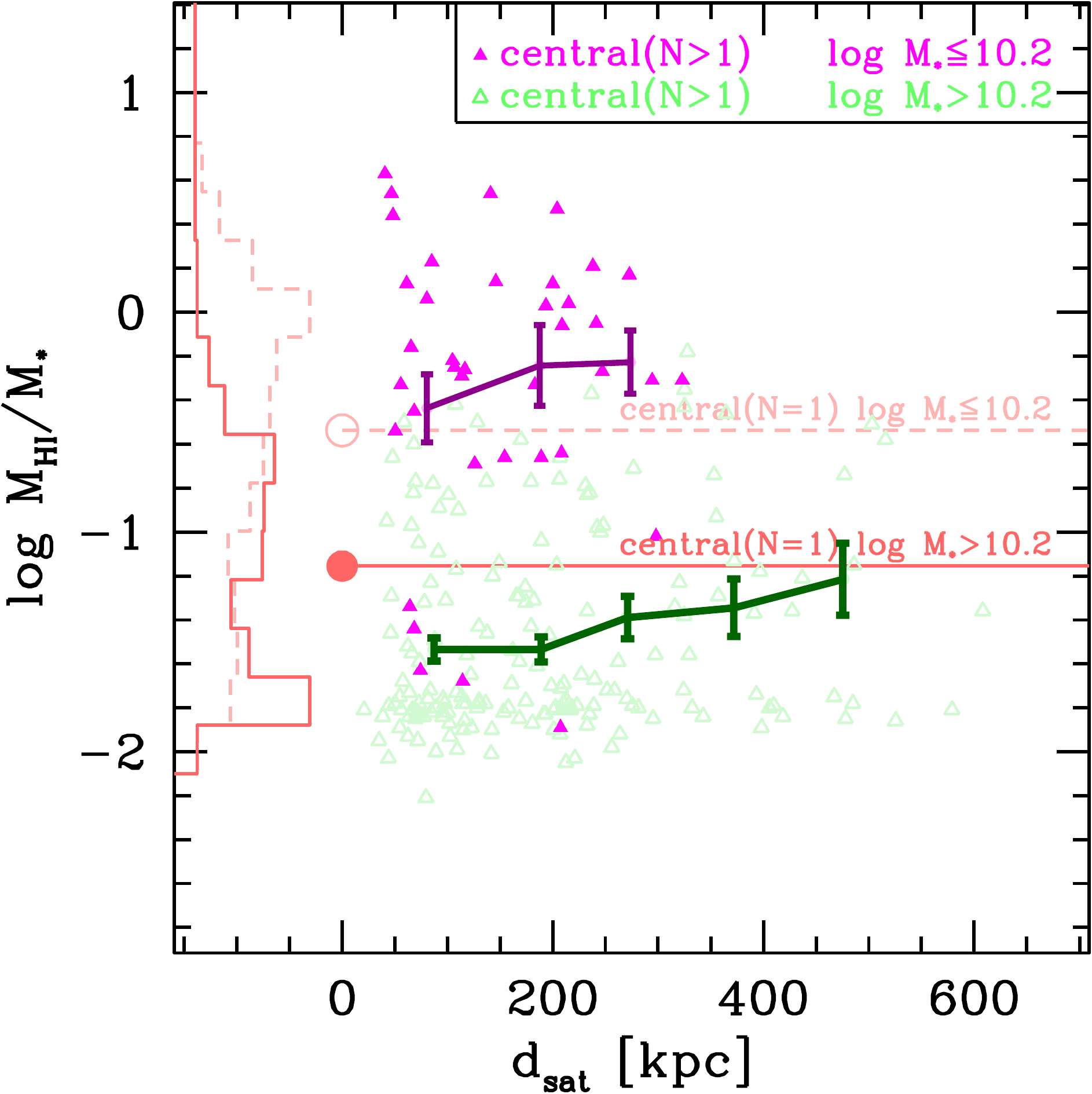}
    \caption{
  \hi \, gas fraction is plotted as a function of projected separation
    between each central galaxy and its nearest satellite galaxy,
    using the same colours and styles as in
    Figure~\ref{fig:Ngal}. Averages and standard errors {of} the mean
    within bins are shown as 
    connected points.
  A decreasing upper envelope is apparent across the full population;
  no strong trends are evident as a function of satellite distance.
    }
    \label{fig:rGF}
\end{figure}

Without {group halo mass estimates for all of the central
  galaxies in our sample, we instead use group multiplicity to compare
  between different groups.
  At a fixed central galaxy stellar mass,
  there is a correlation between group halo mass and group
  multiplicity \citep[for further discussion see Figure~B2
    in][]{han15}. In this sub-section we compare central 
  galaxies in groups of different multiplicity.}

\iffalse
our next best option is to use group multiplicity
as an indicator of the size of the group. There is a general
correlation between halo mass and group multiplicity at fixed stellar
mass, although with 
significant scatter.
 {In this sub-section we include high mass
  group centrals with larger multiplicities (N>4) to increase our
  dynamic range.}
\fi

Figure~\ref{fig:Ngal} shows the \hi \, gas fraction as
a function of group multiplicity (N$_\textrm{gal}$) for the central
galaxies in our sample. The histograms against the y-axis show the \hi
\, gas fractions distributions for the isolated centrals at low
(\Mst/\msun<$10^{10.2}$, in pink
dashed lines) and high (\Mst/\msun$\ge$$10^{10.2}$, as red solid lines)
stellar masses. The average  \hi \, gas 
fractions of these two populations are shown as large dots and
horizontal lines. Our group central galaxies at high (green) and low
(magenta) masses are plotted against their group
multiplicity. Averages within bins of group multiplicity are connected
by thick lines. 

%preplace
\begin{figure}
\centering
\includegraphics[width=0.99\columnwidth]{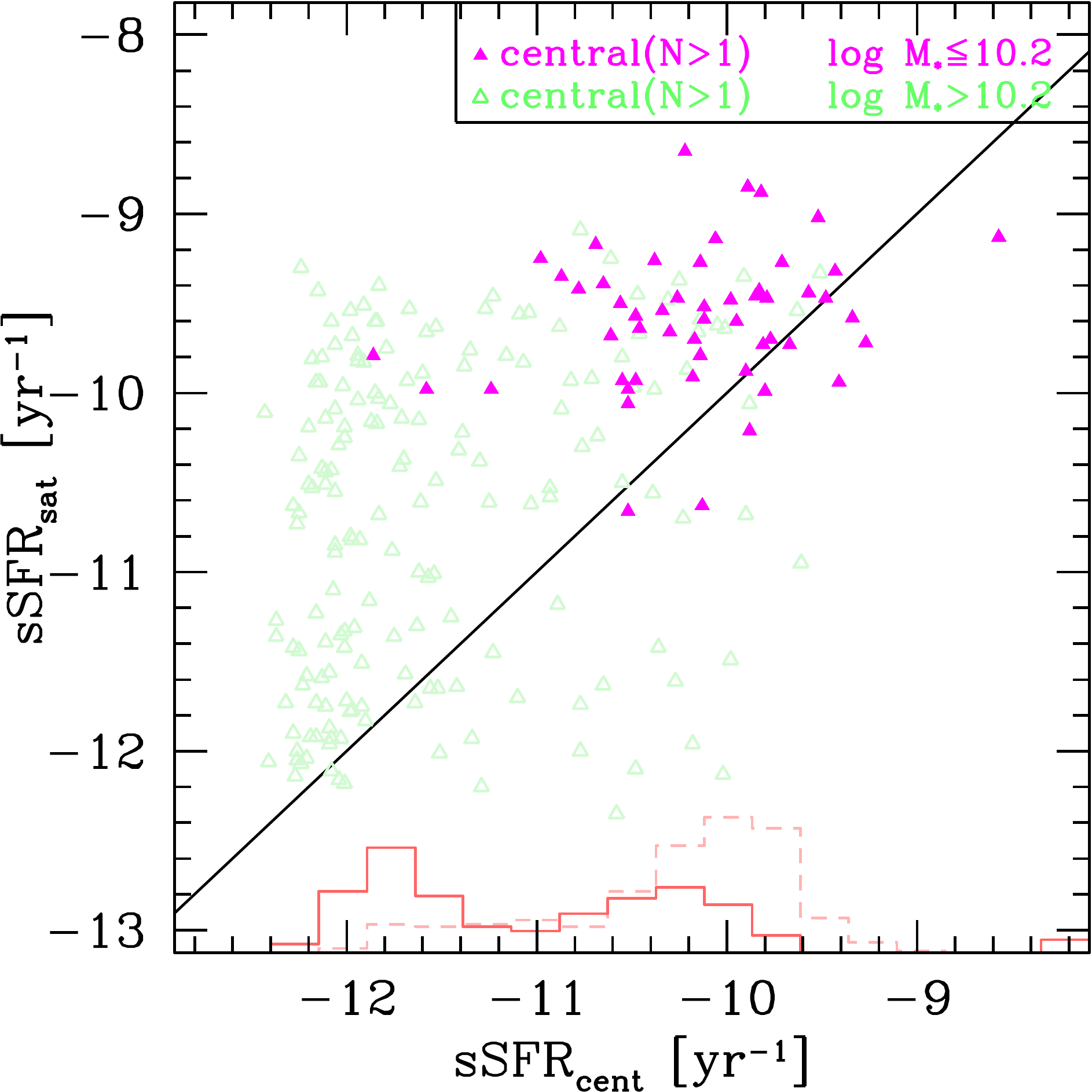}
    \caption{
  The x-axis encodes the sSFR of each group central galaxy in our sample and the
     y-axis shows the sSFR of \textbf{its brightest satellite galaxy},
     which is \textit{not} part of the xGASS sample. The same colour-coding is
     used as in Figure~\ref{fig:Ngal} for high and low mass group
     central galaxies. Also shown on the x-axis are
     histograms of the isolated central galaxy population at low and
     high masses.
  The satellites of the group central galaxies in our sample typically
    have higher sSFRs than their host centrals, and 
    this is especially the case for the low mass group central
    galaxies (shown in purple).
    }
    \label{fig:ssfrssfr}
\end{figure}

The high mass central galaxies (in green) show a continuously
decreasing average \hi \, gas fraction with group multiplicity from
N=1 to 20+, 
such that the most gas-rich high-mass central galaxies are those in
isolation. The low mass central galaxies do not follow this
trend. Instead, the isolated low mass central galaxies have
\textit{lower} average \hi \, gas fractions than those in small groups, and
the most gas-rich galaxies in this population are in groups of
N=2. Similar trends are also evident in sSFR as a
function of {group multiplicity}, but are not plotted here.
Additional explanation is required to
show how a low mass central galaxy with a satellite can be more
\hi-rich and star-forming than an otherwise similar isolated galaxy.

\subsection{Characteristics of these small groups}
\label{sec:character}

Next, we explore the properties of the small groups
that host our low mass central galaxies that are unusually gas
rich and star-forming. We will explore whether they have any other
unusual group properties that could explain the scarcity of
\hi-poor group central galaxies.

First we consider the projected separation between our group central
galaxies and their nearest satellites (d$_\textrm{sat}$) to explore
whether recent or strong interactions from nearby companions may be
responsible for an enhancement in \hi \, and SFR.
Figure~\ref{fig:rGF} shows the \hi \, gas fraction for central galaxies in
our sample as a function of d$_\textrm{sat}$, measured in kpc (in
projection). Central galaxies are binned in two intervals of stellar
mass, as in Figure~\ref{fig:Ngal}. Isolated central galaxies are shown 
as a histogram against the y-axis, separated by mass. 

The \hi \, gas fractions of the group central galaxies at low and high 
masses show no strong trends 
with projected separation to their nearest satellite
galaxy. Best fitting linear relations (not shown) to both the low and
high mass populations yield slopes consistent with zero (low mass
slope is -0.3$\pm$0.7 Mpc$^{-1}$, high mass slope is 0.05$\pm$0.25
Mpc$^{-1}$). We also found no significant trends in sSFR as a function of
d$_\textrm{sat}$ for these populations. The differences in the \hi \,
gas fraction and sSFR in the low mass group central galaxies do not
appear to be strongly dependent on having a nearby companion.

We next consider whether these small groups contain satellite galaxies
that are unique in some way, which could lead to the differences we
observe in the
low mass group central galaxies. This satellite galaxy population is
\textit{not} part of our sample -- these are the satellite members of
the groups \citep[identified in DR7 group catalog of ][]{yang07} that host
the low mass group central galaxies in our 
sample. We have obtained their optical
photometry from the SDSS catalog \citep[Data Release 12,][]{dr12} and
their sSFR values from the DR7 MPA/JHU catalog.

Starting from the population of group central galaxies in our sample,
we identify the brightest satellite galaxy in each group. We
compare the properties of each group central galaxy to its brightest
satellite to see whether there are correlations that help explain
the differences between the group and isolated central
galaxies. In terms of the stellar mass ratio between the brightest
satellite and the group central, we find that {our low mass
  group central galaxies have}
%with N=2-4
 $\langle$log~M$_{*,\text{sat}}/$M$_{*,\text{cent}}\rangle$
= $-$1.09~$\pm$~0.52. The broad distribution of mass ratios shows no trend
with \Mst \, of the group central, with low and high mass group central
galaxies showing a similar range and average mass ratios (about
10\%). From this we can conclude that our population of low mass group
central galaxies are not members of groups with unusual mass ratios,
but appear consistent with typical values. Larger groups (N>4) have a
narrower range of less extreme mass 
ratios with an average of
$\langle$log~M$_{*,\text{sat}}/$M$_{*,\text{cent}}\rangle$~=~$-0.40$~$\pm$~0.28
(all of these large groups  
have central galaxies with \Mst/\msun$\gtrsim$10$^{11}$).

%mass ratios:
%overall group centrals (N=2-4)
%   <log M2/M1> = -1.09 +/- 0.52
%  low mass 2-4
%     <log M2/M1> = -1.13 +/- 0.49
%  high mass 2-4
%     <log M2/M1> = -1.08 +/- 0.53
%group centrals with N>4: (high mass)
%  <log M2/M1> = -0.40 +/- 0.28

We also compare sSFR values between group central galaxies and their
brightest satellites in Figure~\ref{fig:ssfrssfr}. Here the x-axis
shows the sSFR of our group central galaxies and the y-axis shows the
sSFR of their brightest satellite galaxies. As {the stellar
  masses of satellites, by definition, are less than the
  stellar mass of their central galaxies}, and lower 
mass galaxies typically have higher sSFRs, most of the points lie
above the unity line (meaning that most satellite galaxies have a  
larger sSFR than their host central galaxies).

%dropped now:
%%preplace
%\begin{figure*}
%\centering
%\includegraphics[width=1.3\columnwidth]{plot_aden_GF_envB_M_3-crop.pdf}\\
%\includegraphics[width=1.3\columnwidth]{plot_aden_SSFR_envB_M_3-crop.pdf}
%    \caption{
% Top panel shows \hi \, gas fraction and bottom panel shows sSFR, both
% as functions of 1-Mpc environmental density for central galaxies in
% our sample.
% Isolated central galaxies are shown at low (red line, grey circles)
% and high masses (pink line, grey dots), and group central galaxies
% are shown at low (magenta line, filled triangles) and high masses
% (green line, open triangles).
% %Color-coding is as in Figure~\ref{fig:Ngal}, but with isolated
% %central galaxies at low mass shown with grey circles (and averages
% %connected by a red line) and at high mass with grey dots (and a pink
% %line).
%   Averages are only shown in bins with at least five galaxies.
%  The average \hi \, gas fractions show a weak trend to decrease at higher
%  densities, and the sSFRs do not show any significant 
%  trends. At fixed density, low mass group
%  central galaxies (magenta line) are on average more \hi-rich and
%  star-forming than low mass isolated galaxies (red line).
%    }
%    \label{fig:den}
%\end{figure*}

%preplace
\begin{figure*}
\centering
\includegraphics[width=0.90\textwidth]{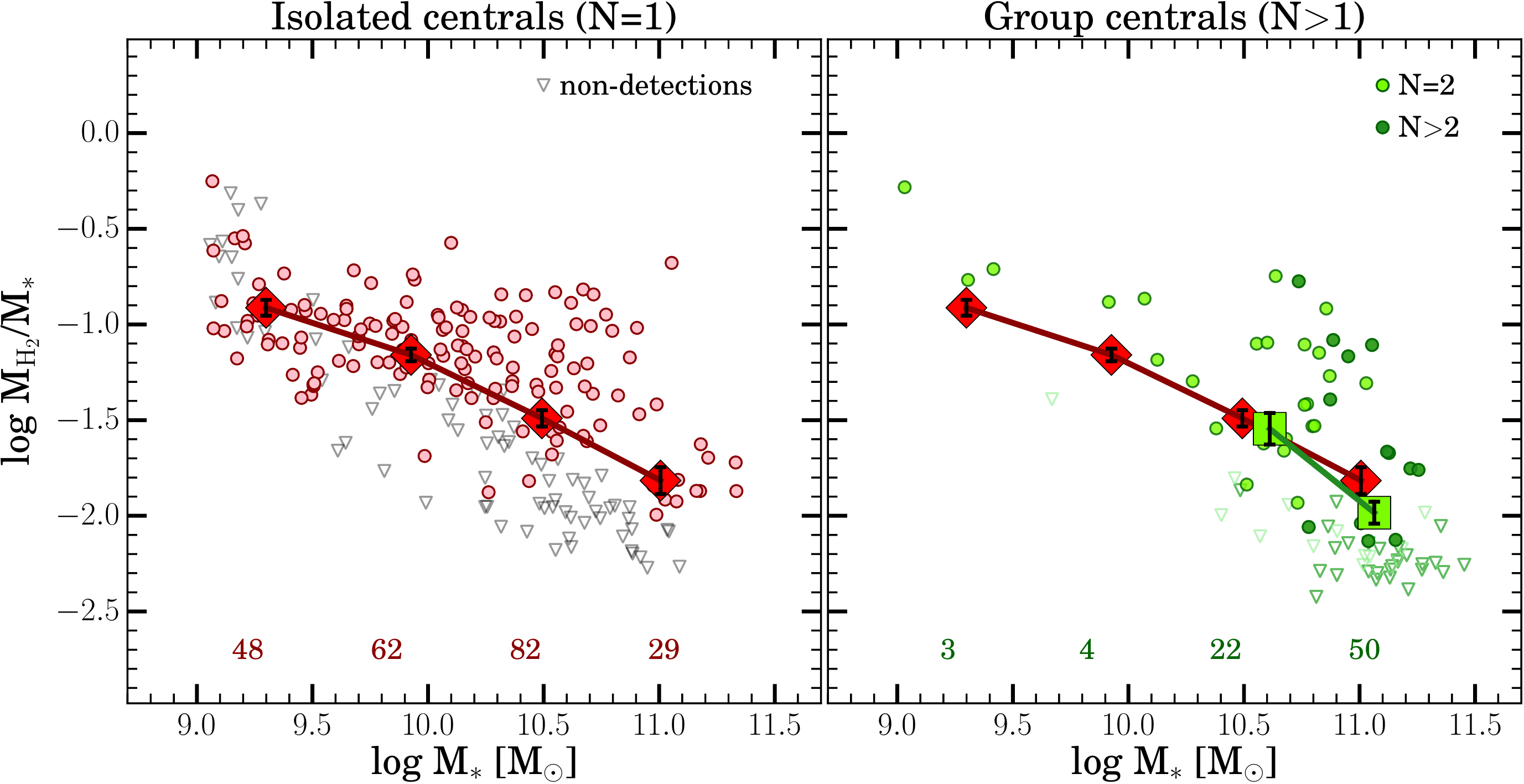}
    \caption{
   \H2 gas fraction of central galaxies is shown as a function of
   stellar mass, for isolated (left panel) and group (right panel)
   environments. The same colours and averaging are used as in
     Figure~\ref{fig:envHI}. Large {symbols} show averages within bins
     of stellar mass when there are more than 10 points in the bin.
     While there are only 7 group central galaxies at low stellar
     mass (not enough for meaningful averages), these results are not
     inconsistent with the \hi \, trends.
    }
    \label{fig:envH2}
\end{figure*}

The satellites of the low mass group central galaxies appear to be
consistent with this trend, and have higher sSFRs than their
centrals in all but a few cases. Of the high mass group central
galaxies, $\sim$85\% have satellites with higher sSFRs. Similarly,
$\sim$80\% of low mass group central galaxies have a satellite with a
higher sSFR. The satellite galaxies of the low mass central
galaxies have some of the highest sSFRs of any (brightest) satellites
in this sample, which is not surprising as they are also some of the
lowest mass satellites in this sample. This trend indicates that the
low mass group central galaxies have not become star-forming (or, by
extension, gas-rich) at the expense of their satellites.

Finally, we also consider the {local} environmental density around
each of the central galaxies in our sample, measured in 1-Mpc
apertures (see Section~\ref{sec:env}). This type of density estimate
represents a different environmental metric than isolated vs. group
categorization. Instead, the 1-Mpc scale density estimate relates
to the nearby environment. This density measurement is not
dependent on group-finding algorithms or any parametrization of large
scale structure, and instead simply measures the number of galaxies
with \Mst/\msun$\ge$$10^9$ in the vicinity of each galaxy in our sample.

We find no significant trends in \hi \, gas fraction or sSFR as a
function of local density. {For central galaxies in our
  sample, these 1-Mpc local density
  measurements span two orders of magnitude, from 0.3 to 50 galaxies
  per Mpc$^2$.} 
We note that the average differences in \hi
\, and sSFR between group and isolated central galaxies are still
observed at fixed value of local density. That is, we find both group
and isolated central galaxies distributed across all values of local
density, and neither \hi \, gas fraction nor sSFR shows a significant
dependence on density.

\subsection{\H2 content of low mass group central galaxies}
\label{sec:h2}

While molecular gas observations are significantly more difficult to
obtain than 21cm atomic hydrogen observations, the \H2 
properties of galaxies are critical to their star formation and
evolutionary processes. Although \hi \, is the fuel for star formation in
galaxies, the detailed process of forming stars occurs in pockets of
cold molecular gas \citep{kennicuttevans12}, and molecular gas is
observed to correlate much better with star formation than \hi \,
\citep[][ and references therein]{leroy13}. Given this direct linkage
between \H2 and star formation, we explore whether the \H2
observations of low mass group central galaxies show similar
differences as seen in \hi \, and SFR.

Figure~\ref{fig:envH2} shows the \H2 gas fraction vs. stellar mass for
group central galaxies separated by group multiplicity (compare
Figure~\ref{fig:envHI}). While it is immediately apparent that we have
fairly poor statistics (less than a third of our sample has been
observed in \H2), it is still interesting to compare the
findings. Even with only seven low mass group central galaxies
observed in \H2, 6/7 have higher \H2 gas fractions than the average of
isolated galaxies at the same stellar masses. While this difference is
only weakly significant, it is not inconsistent with our findings in
\hi \, and SFR. It suggests that the population of low mass central
galaxies in groups may have, on average, more molecular gas than those
in isolation, as expected given that their average \hi \, gas
fractions and sSFRs are also larger.

\section{Discussion}
\label{sec:discussion}

While previous works have shown that galaxies in groups have larger
average star formation rates over those in isolation at fixed stellar
mass, this is the first study to 
show that there is also a difference in the \hi \, content of galaxies
in small groups as well. We have shown that low mass
($10^{9}$$\le$\Mst/\msun<$10^{10.2}$) central galaxies in small
{(mostly N=2)} groups
have larger average \hi \, fractions compared to isolated central
galaxies. This difference is driven moreso by a relative lack of
gas-poor group central galaxies, rather than by an enhancement in
group central galaxies themselves. In the gas-rich star-forming
regime, isolated and group central galaxies have similar \hi \, and
sSFR properties; it is largely in the gas-poor passive regime where
group and isolated central galaxies differ. We next discuss and 
interpret this result and its place within the context of 
other studies of gas in galaxies across different environments.

Broadly, this difference in the \hi \, properties of group central
galaxies is consistent 
with the interpretation of the findings of \citet{moorman14} and
\citet{jones16}, who use the ALFALFA survey to measure
the \hi \, mass function in different environments. Both groups found that
galaxies located in higher density regions (e.g., walls, filaments,
groups) have $0.1$-$0.2$~dex larger \hi \, masses than those in low density
regions (e.g., voids). Even though they only probe the gas-rich
regime and are not comparing galaxies at fixed stellar mass, there
is broad agreement between these ALFALFA-based \hi \, environmental
studies and our deeper observations of central galaxies.

\subsection{Connections to galaxy pairs and conformity}
\label{sec:pairs}

The differences we find in gas fraction and sSFR in low mass
group central galaxies are similar to enhancements observed in pairs
of galaxies. Galaxy pairs are typically defined to have projected
separations less than 100~kpc and velocity differences <350~\kms \,
\citep{lambas03}, with some studies of interacting pairs reaching
even smaller separations \citep[e.g.,][]{wong11}. 
When compared with otherwise-similar non-paired galaxies, the SFR in
galaxy pair 
members is typically enhanced by 10-20\% in pairs with up to 150~kpc
separations \citep{patton13}, but with the strongest
enhancements (up to 30\%) at the smallest separations \citep{wong11}.
{Local environment} density also plays a role, as isolated pairs
have greater 
SFR enhancements compared to those in higher density regions
\citep{ellison10, wong11}.

In our sample, {$\sim$30\%} of our low mass group central %16/55
galaxies could be 
classified as galaxy pairs (i.e., they have only one satellite which is
within a projected distance of 100~kpc and 350~\kms \, of
velocity). As we have removed all sources with significant
\hi-confusion (see Section~\ref{sec:sample}), we have effectively
removed the closest pairs from our sample. {These ``pair
centrals'' in our sample appear to be similar to 
the other low mass group central galaxies,} with an average
satellite-to-central mass ratio of $\sim$10\% and following the same
distribution of \hi \, gas fraction 
(three of the five most \hi-poor low mass group centrals are in this
category). {Broadening the definition of pairs to include
  galaxies with separations up to 200~kpc would include $\sim$65\% of our %36/55
  low mass group central galaxies. As was the case for the 100~kpc
  pairs, these wider pairs are also similar to the full sample, and have larger
  average \hi \, gas fractions than isolated central galaxies.}

This is consistent with Figure~\ref{fig:rGF}, which shows
that nearby satellite galaxies do not have a significant effect on the
\hi \, content of central galaxies in our sample. While the pairs in
our sample do have higher average \hi \, gas fractions, the low mass
group central galaxies of multiplicity N=2 with larger separations
(spatially or kinematically) are also more \hi-rich on average. The
{enhancements in \hi \, seen in previous studies of galaxies
  pairs are} not enough to explain {the \hi \, properties} of
our population of group central galaxies.

Simulations of galaxies in pairs have also shown that their SFR is
statistically enhanced compared with un-paired galaxies, and that the
SFR enhancement depends sensitively on the orbital parameters of the
galaxy-galaxy interaction \citep{perez06}. The tidal and
hydrodynamical interactions are the triggers for star formation and
increase the 
gas consumption rate \citep{park2008}. More recently,
cosmological hydrodynamical simulations have found that galaxies in
pairs also have higher \hi \, gas fractions than similar 
un-paired galaxies \citep{tonnesen12}. 
This gas enhancement is thought
to be a result of gravitationally-induced hydrodynamical effects that
increase cold gas formation from the hot halo. Observations of low
metallicity gas in the inner parts of the disks of galaxy pairs also
suggest that interactions may trigger inflows of metal-poor gas from
the halo \citep{rampazzo05, kewley10}. These galaxy-pair interactions
may be analogous to some of the early stages of galaxy pre-processing,
and are an important component of group evolution.

When comparing the sSFRs of our group central galaxies with their
brightest satellites (shown in Figure~\ref{fig:ssfrssfr}), the low
mass population was not distinctive from the high mass; the 13
``pair centrals'' among these are also not unique or extreme. It
appears that the groups hosting star-forming low mass central galaxies
also host star-forming satellite galaxies. Optical studies have found
similar conformity in the optical colours of galaxies in groups and
clusters \citep{kauffmann10}. Recently, \hi \, studies have also found
that \hi-rich central galaxies are more
likely to be found in \hi-rich environments \citep{wang15}. 
Combined, our findings and these results suggest that the satellite
galaxies in these small groups are likely themselves gas-rich, and
have not been stripped in order to enhance their group central
galaxies. Rather, our gas-rich low mass group central galaxies
likely live in groups that are also gas-rich, presumably as a result
of their {local} environment.

\subsection{The evolution of low mass central galaxies}
\label{sec:cent}

The stellar mass of a central galaxy and of its total halo are
considered inter-related fundamental parameters that control the
evolution of galaxies in groups. However, at low stellar masses and in
small groups, some of the relations from larger galaxies break
down. For example, central stellar mass is a good tracer of halo and
clustering properties for only the most massive central galaxies
\citep[\Mst/\msun$>$$10^{11}$,][]{wang16}. For 
central galaxies in large halos (M$_h$$\sim$$10^{13}$\msun), most of
their stellar mass growth comes from mergers, while in smaller halos
(M$_h$/\msun$\sim$$10^{11.3}$) they grow 
primarily through star formation \citep{zehavi12}. Furthermore, the
relationship between central galaxy stellar mass growth and halo mass
growth shows a dependence on environmental density
\citep{tonnesen15}. At lower halo 
masses (i.e., in small groups) and lower stellar mass, the evolution of
central galaxies depends most strongly on secular factors, such as the
availability of gas and presence of star formation.

Low mass central galaxies in small groups are only weakly feeling the
quenching effects of their group environment, and are still strongly
affected by their own internal evolutionary processes. With halo
masses too low to 
quench star formation \citep[e.g.,][and references therein]{yang13, zu16},
these central galaxies are 
at an intermediate evolutionary stage between field and cluster
environments. 
{In groups this size and at these stellar masses, our low
  mass group central galaxies are unlikely to host an AGN \citep[see,
    e.g., Table~1 of ][]{ellison08}}, and their star formation is not
strong enough to remove gas through galactic winds. Between their mild
environments and 
lack of central feedback, they are overall unlikely to rapidly lose
their gas and will continue forming stars.

To better understand the possible evolutionary paths that allow these
low mass group central galaxies to remain more \hi-rich than their
isolated counterparts, we 
next look at 
previous results from observations and simulations on the role of gas
in galaxies across different environments.

\subsection{Observing and modeling gas in groups}

Using \hi \, observations of 72 compact groups,
\citet{verdes-montenegro2001} proposed an evolutionary
sequence for gas in groups of galaxies (see their Figure~7). Their
sequence begins with a 
{compact} group of a small number of 
mostly late-type gas-rich star-forming member galaxies {which
  have a low level of interaction}. Next,
galaxy-galaxy interactions produce tidal tails and remove or
redistribute the gas from the galaxies, and morphologically transform
them as well. Finally, most of the galaxies have been transformed into
gas-poor early types and most of the \hi \, left in the group is in the
form of a hot halo.
The groups that host the low mass group central galaxies in our sample
are comparable to groups at the beginning of this evolutionary
sequence: their central and satellite galaxies still have ongoing star
formation, and the central galaxies have significant amounts of cold
gas. However, it is not clear when or if our group central galaxies
will evolve into gas-poor early-types, as the members of these small
groups are more spread out (and have longer dynamical times) than the
compact groups of \citet{verdes-montenegro2001}.

On larger scales, gravitational and hydrodynamical simulations can
provide insight on the origin and evolution of these groups. According
to structure formation theory, dark matter halos 
above a certain mass collapse \citep{birnboimdekel2003}, material from
their surroundings tends to form filaments which funnel it onto the
centers of the halos. This filamentary assembly is evident in dark
matter-only N-body simulations \citep[e.g.,][]{aubert04} as well as in
cosmological hydrodynamical simulations \citep[e.g.,][]{pichon11},
{in which both satellite galaxies and cold flows of gas follow
  filamentary structures as halos grow in the high-redshift
  universe. At lower redshifts, satellite galaxies similarly trace the
  filamentary structures \citep{welker16}, and these inflows are
  continuous from cosmological scales down to galactic 
scales \citep{danovich15}.}

The large scale flows into galaxies and groups can be either cold gas
or hot gas \citep[e.g.,][]{ocvirk08}. For galaxies with
\Mst/\msun$>10^{10.3}$ and in groups of 
M$_\textrm{h}$/\msun$>10^{11.4}$, most of the accretion is in the form of hot
gas \citep{keres05}. However, for the low mass galaxies in smaller
groups, accretion along filaments provides the primary
source of fuel for star formation in the low redshift universe
\citep{brooks09}. Minor mergers of \hi-rich galaxies also contribute
significantly to the growth of low mass galaxies \citep{lehnert16}.

Most relevant to this work is the strong connection in simulations
between group environments and filamentary structure in the cosmic
web. It is likely that our low mass group central galaxies are being
fed by the filamentary structures in which they are
embedded. Galaxy groups form at the intersections of these
{filaments. The groups grow as galaxies and gas fall in along
  the filaments and can feed the group central galaxies, making them
  more likely to be gas-rich than those in isolation.}

At fixed stellar mass, a central
galaxy in a group will have access to more gas than an isolated
central galaxy. At higher stellar masses and in larger groups there
are various quenching mechanisms \citep[e.g.,][and references
  therein]{yang13,zu16} which reduce the amount of cold gas 
in central galaxies, erasing any evidence of this
enhancement. However, at low stellar masses and in small groups, this
\hi \, richness can persist in central galaxies and is observed for the
first time in this work.

\section{Summary}
\label{sec:summary}

We use a sample of central galaxies in groups and isolation to
investigate the effects of environment on their cold gas and
star-formation properties. In particular, we find that low mass
($10^{9}$$\le$\Mst/\msun$<$$10^{10.2}$) group central galaxies have
gas fractions and 
sSFRs that are larger than isolated central galaxies by
0.2-0.3~dex, at fixed stellar mass. This difference is driven largely
by the gas-poor central galaxies, which are found significantly more
often in isolation than in groups. The distinction between group and
isolated central galaxies is
consistently found across multiple group catalogs, our \hi \, stacking
analysis, and in larger samples of galaxies. These low mass central
galaxies are found in small groups ({usually N=2}) whose satellite members also
have larger sSFRs. {As discussed in Section~\ref{sec:env},
  identifying the central galaxy in small groups is difficult; in this
  work we simply define central galaxies as the most massive member
  of a group.}

These small, gas-rich, star-forming groups are found in moderately
over-dense environments (intermediate between isolation and clusters)
and might represent an early stage of group 
evolution. Their low mass central galaxies have a large \hi \, gas
reservoir, which {simulations suggest} is likely fed by gas
infall along filaments or from 
earlier  mergers of gas-rich satellites. Low mass central galaxies in
small groups have not yet grown large enough to experience significant
environmental or internal quenching. As
these groups continue to grow through star formation and
mergers, their central galaxies will become less gas rich.

Further work is needed using much larger samples of galaxies across 
environments and at moderate and low \hi \, gas fractions. Note that
without reaching very low \hi \, gas fractions (and upper limits), the
populations of central, isolated, and satellite galaxies would have
been indistinguishable. While time-consuming, these deep \hi \,
observations are vital for understanding the evolutionary path of
galaxies between gas-rich field and gas-poor cluster environments.

\section*{Acknowledgements}

%The Acknowledgements section is not numbered. Here you can thank helpful
%colleagues, acknowledge funding agencies, telescopes and facilities used etc.
%Try to keep it short.

We thank our anonymous referee for very helpful suggestions which have
improved this manuscript.
We thank Charlotte Welker and Katinka Ger\'eb for helpful
discussions.

SJ, BC, and LC acknowledge support from the Australian Research
Council's Discovery Project funding scheme (DP150101734).
BC is the recipient of an Australian Research Council Future
Fellowship (FT120100660). 
AS acknowledges the support of the Royal Society through the award of
a University Research Fellowship.

This research has made use of NASA's Astrophysics Data System, and
also the 
NASA/IPAC Extragalactic Database (NED), which is operated by the Jet
Propulsion Laboratory, California Institute of Technology, under
contract with the National Aeronautics and Space Administration. This
research has also made extensive use of the invaluable Tool for
OPerations on  Catalogues And Tables
\citep[TOPCAT\footnote{\url{http://www.starlink.ac.uk/topcat/}}, 
][]{taylor05}.

Some of the data presented in this paper were obtained from the
Mikulski Archive for Space Telescopes (MAST). STScI is operated by the
Association of Universities for Research in Astronomy, Inc., under
NASA contract NAS5-26555. Support for MAST for non-HST data is
provided by the NASA Office of Space Science via grant NNX09AF08G and
by other grants and contracts.

Funding for SDSS-III has been provided by the Alfred P. Sloan
Foundation, the Participating Institutions, the National Science
Foundation, and the U.S. Department of Energy Office of Science. The
SDSS-III web site is \url{http://www.sdss3.org/}.

SDSS-III is managed by the Astrophysical Research Consortium for the
Participating Institutions of the SDSS-III Collaboration including the
University of Arizona, the Brazilian Participation Group, Brookhaven
National Laboratory, Carnegie Mellon University, University of
Florida, the French Participation Group, the German Participation
Group, Harvard University, the Instituto de Astrofisica de Canarias,
the Michigan State/Notre Dame/JINA Participation Group, Johns Hopkins
University, Lawrence Berkeley National Laboratory, Max Planck
Institute for Astrophysics, Max Planck Institute for Extraterrestrial
Physics, New Mexico State University, New York University, Ohio State
University, Pennsylvania State University, University of Portsmouth,
Princeton University, the Spanish Participation Group, University of
Tokyo, University of Utah, Vanderbilt University, University of
Virginia, University of Washington, and Yale University.

%%%%%%%%%%%%%%%%%%%%%%%%%%%%%%%%%%%%%%%%%%%%%%%%%%

%%%%%%%%%%%%%%%%%%%% REFERENCES %%%%%%%%%%%%%%%%%%

%%%%%%%%%%%%%%%%%%%%%%%%%%%%%%%%%%%%%%%%%%%%%%%%%%

%%%%%%%%%%%%%%%%% APPENDICES %%%%%%%%%%%%%%%%%%%%%

\appendix

\section{Modifications to the DR7 Yang \etal group catalog}
\label{sec:yangmod}

\begin{table*}
  \centering
	\caption{NYU IDs and other details for the cases of shredding among non-confused central galaxies
          from the ``Group B'' DR7 group catalog of
          \citet{yang07}.}
	\label{tab:yangmod2}
	\begin{tabular}{lllllllll} 
          \hline
          xGASS & NYU ID & env.  & env.  & Group ID & NYU ID(s)  & N$_\text{gal}$ & N$_\text{gal}$ & comment \\
                & target & orig. & corr. &          & removed    & orig.         & corr.        & \\
          (1) & (2) & (3) & (4) & (5) & (6) & (7) & (8) & (9) \\
          \hline
%these are from killid_B 
%hidesat          111080 &     sat &     sat &   900 & 2395553 & 17 & 16 & xGASS galaxy shredded \\
          109081 & 2483120 &   group &   group &  6743 & 2483121, 2483122 &  5 &  3 & target is shredded into 3 components  \\
%hidesat          112080 &     sat &     sat &  9795 & 2249918 &  4 &  4 & another member shredded \\
%hidesat            4132 &     sat &     sat & 10448 &  150388 &  3 &  2 & xGASS~4132 shredded into 2 components \\
            3917 & 150390 &   group &   group & 10448 &  150388 &  3 &  2 & target is shredded into 2 components \\
%%        114146 &   group &   iso?? & 15369 & 1422661 &  3 &  2 & actually dropped - did not match source \\
%hidesat          114015 &     sat &     sat & 16886 & 2008714 &  3 &  2 & group central is shredded into 2 components \\
          123010 & 348076 &   group & isol. & 22677 &  348078 &  2 &  1 & target is shredded - should be isolated \\
%         114110 &     sat & iso???  & 26664 &  598943 &  2 &  2 & xgass is confused anyway. bad
          109065 & 948834 &   group & isol. & 30433 &  948833 &  2 &  1 & \textquotesingle \textquotesingle \\ %target is shredded - should be isolated \\
           23070 & 1073305 &   group & isol. & 32293 & 1073306 &  2 &  1 & \textquotesingle \textquotesingle \\ %target is shredded - should be isolated \\
          %NYUID= 1223869  (merger) apparently some other galaxy that was shredded. not in xGASS anymore
          %NYUID= 1388046  (merger) apparently some other galaxy that was shredded. not in xGASS anymore
           39346 & 1423425 &   group & isol. & 39398 & 1423426 &  2 &  1 & \textquotesingle \textquotesingle \\ %target is shredded - should be isolated\\
          110013 & 1821976 &   group & isol. & 41473 & 1821977 &  2 &  1 & \textquotesingle \textquotesingle \\ %target is shredded - should be isolated\\
%         xGASS 114041 &   NYUID = 1447971  that is the xGASS target, and it is removed. confused anyway too
%%          108029 &     sat & isol. & 32412 & 1088574 &  2 &  1 & merging pair with mis-identified SDSS spectra, but is confused as well so dropped\\
%not in RS          113151 &  265576 &   group & isol. & 21155 &  265577 &  2 &  1 & \textquotesingle \textquotesingle \\ %target is shredded - should be isolated\\
%NOT in RS          113119 & 2276996 &   group & isol. & 48225 & 2276997 &  2 &  1 & \textquotesingle \textquotesingle \\ %target is shredded - should be isolated\\
%not in RS          110009 & 2135746 &   group & isol. & 46274 & 2135747 &  2 &  1 & \textquotesingle \textquotesingle \\ %target is shredded - should be isolated\\
%next, check killid_C
          %2371488  not in C
          %1932693  not in C
          %2483121  already mentioned above
          %2483122  already mentioned above
          %948835   not in C
          %948833   already mentioned above
          %446796   not in C
          %1307831  not in C
          %1821977  already mentioned above
          \hline
          \multicolumn{9}{l}{
          \text{1) xGASS ID; 
            2) NYU~ID of xGASS target;
            3) original environmental identity of xGASS target;
            4) corrected environmental }}\\
          \multicolumn{9}{l}{
          \text{
            identity;
            5) group ID in \citet{yang07} DR7 B catalog;
            6) NYU~ID of removed shredded component;
            7) original  }}\\
          \multicolumn{9}{l}{
          \text{
            group multiplicity of xGASS target;
            8) corrected group multiplicity;
            9) detailed description.  }}
	\end{tabular}
\end{table*}

As described in Section~\ref{sec:env}, we have identified cases of
galaxy shredding and ``false pairs'' in the DR7 \citet{yang07} group
catalog. To correct these, we have removed the smaller shredded
component and retained only the main galaxy in the catalog (and have
re-computed all relevant group multiplicities and satellite separations).

%As an example, we discuss GASS~4132. This target is a spiral galaxy
%identified as a satellite galaxy of a nearby central galaxy (i.e., a group of
%N=2). However, GASS~4132 (the satellite spiral galaxy) is 
%shredded into two sources in all of the \citet{yang07} DR7 catalogs
%(``A'', ``B'', and ``C''), so that the group incorrectly has
%3 members. We remove the extraneous source (with NYU~ID=150388),
%correct the group multiplicity to N=2, and re-compute the relevant
%separation between the central and satellite galaxies in this group.

We closely examine all of our central galaxies which are not confused
in \hi \, and identify those that are affected by shredding. For
central galaxies in groups, we also examine their satellite galaxy
members to identify any which have been shredded (leading to an
inflated group multiplicity). When we find that galaxies have been
shredded we remove the shredded component from the catalog (based on
its NYU~ID) and re-determine environmental identities and group
multiplicities. Complete details about our corrections are listed in
Table~\ref{tab:yangmod2}.

In two cases (xGASS~109081 and xGASS~3917) our target is a group
central galaxy and is shredded, which incorrectly increases its group
multiplicity (by 2 and 1 galaxies, respectively). We remove the
shredded components and re-compute the 
group multiplicities. In {the remaining} cases our target galaxy
is shredded 
into two sources, one of which is identified as a ``group
central'' and the other as a ``satellite''. We remove the shredded
component, correct the group multiplicity to N=1, and change the
environmental identity from ``group central'' to ``isolated central''
galaxy. In no cases were the satellites around (non-confused) xGASS
group central galaxies shredded.

\section{Results with different subsets/datasets}
\label{sec:others}

The following paragraphs and figures describe and show confirmations
of our main results using different subsets of our galaxies or
different group catalogs. 

\begin{figure}
\centering
\includegraphics[width=.99\columnwidth]{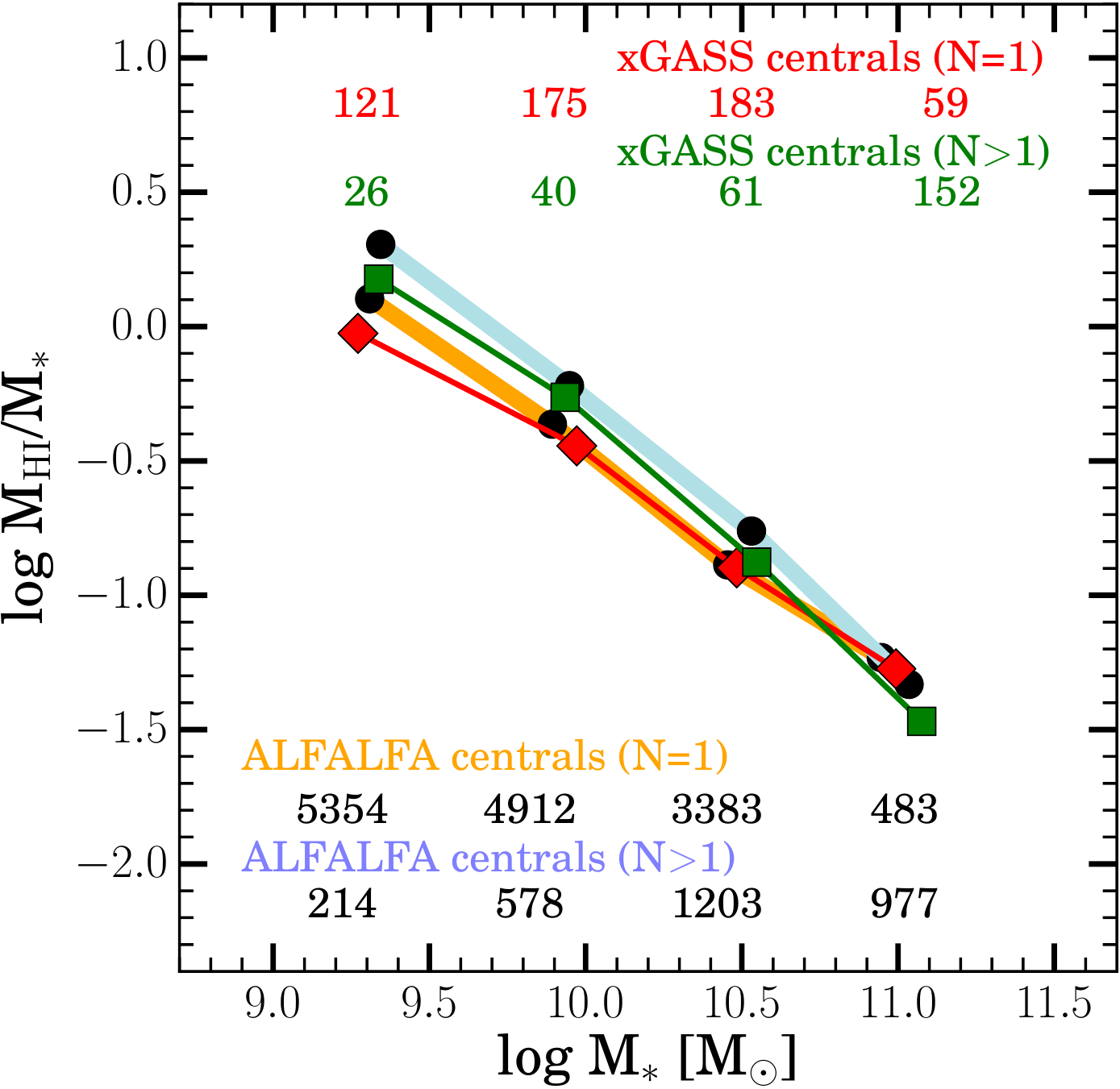}\\
    \caption{
      Same as Figure~\ref{fig:envHIstack}, including all
      \hi-confused galaxies from xGASS and the ALFALFA
      stacking sample. 
      {
        Green/red lines and dots show stacked values of \hi \, gas
          fraction in bins of the xGASS central galaxies in groups
          (N>1) and isolation (N=1);
        blue/yellow lines and black dots show the same for the ALFALFA
        stacking sample.
      Our main results are unchanged when including \hi-confused galaxies.}
    }
    \label{fig:stackconf}
\end{figure}

{
First, Figure~\ref{fig:stackconf} shows the stacked \hi \, gas fraction
scaling relations (compare Figure~\ref{fig:envHIstack}) for our sample
including confused objects and for the ALFALFA stacking sample (also
including confused sources). \added{Note that the average \hi \, gas
  fractions of the xGASS group centrals (in green) increase
  by up to $\sim$0.15~dex relative to their values in
  Figure~\ref{fig:envHIstack}, as additional gas is being included
  from nearby confused satellites. Overall, }
 there is no significant difference from
the stacking comparison with confused galaxies removed, and our main
conclusions remain unchanged.}

\begin{figure*}
\centering
\includegraphics[width=0.99\textwidth]{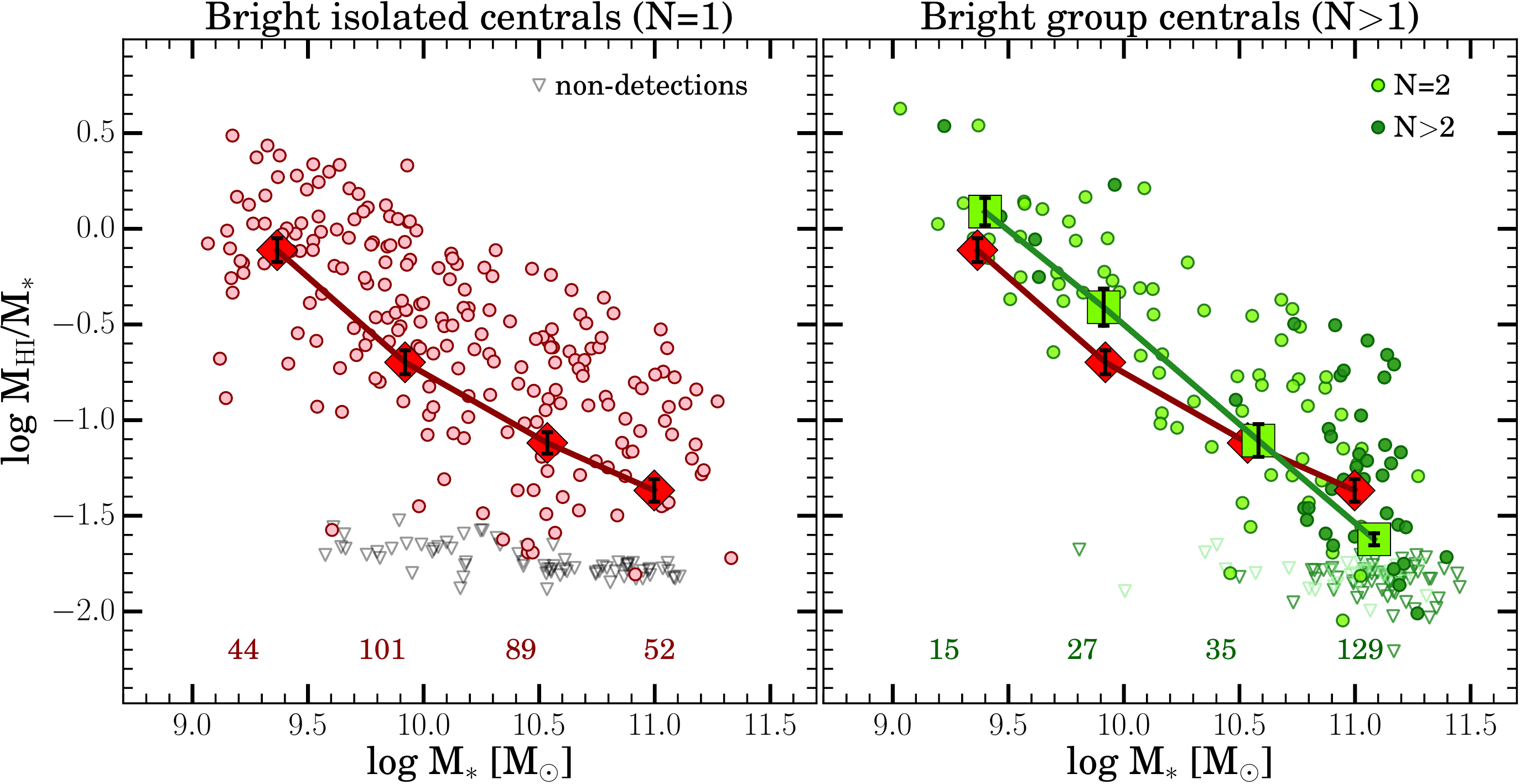}\\
\vspace{0.5cm}
\includegraphics[width=0.99\textwidth]{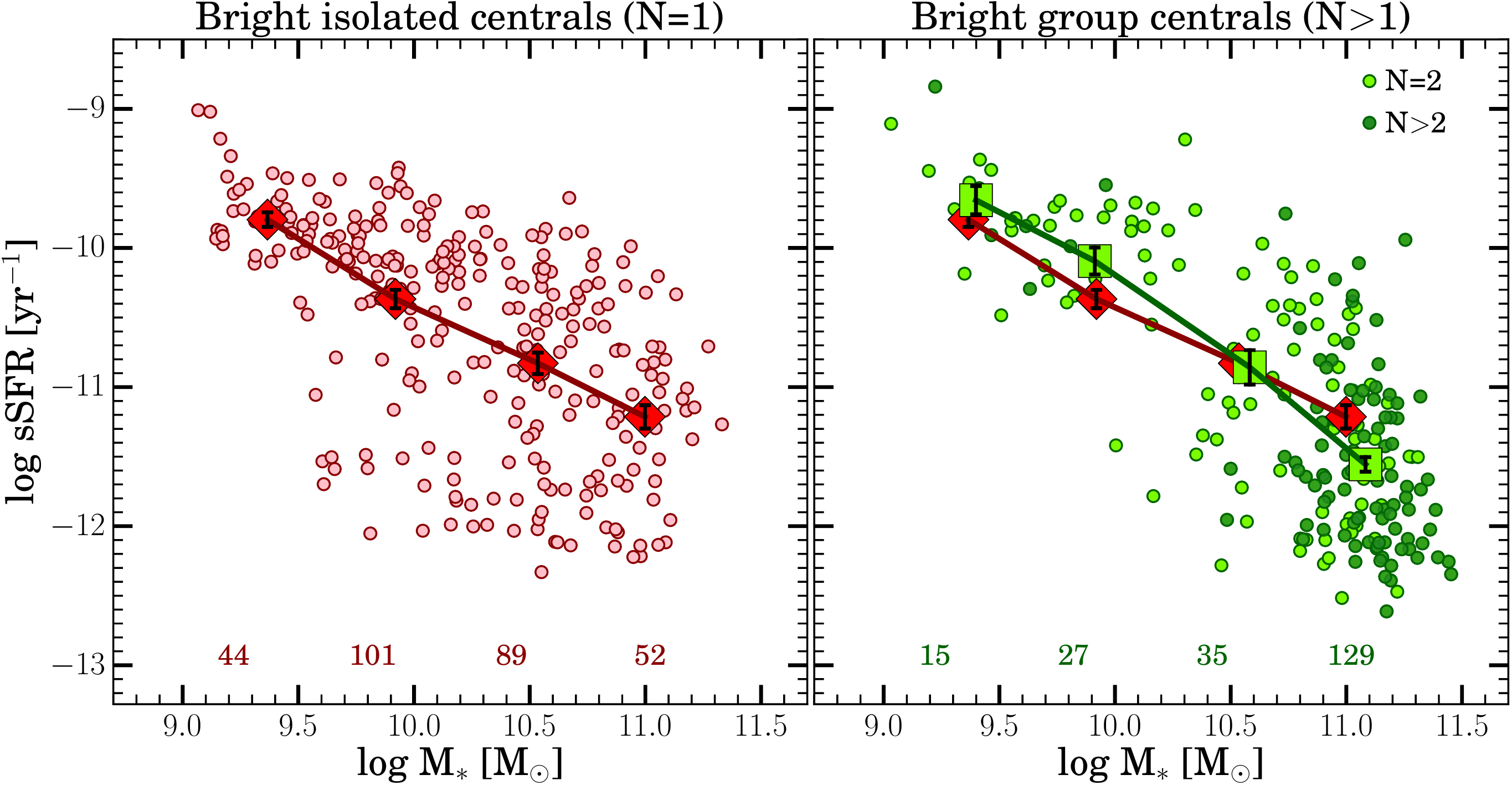}
    \caption{
   Same as Figures~\ref{fig:envHI} and \ref{fig:sfr}, but for only the
   ``bright'' subset of central galaxies, {at least 2.5~mag
     brighter than the SDSS spectroscopic survey limit}. Our main
   results are    unchanged.
    }
    \label{fig:bright_GFSF}
\end{figure*}

Next, Figure~\ref{fig:bright_GFSF} shows our main \hi \, gas fraction and
sSFR relations using only the ``bright'' subset of central galaxies,
as described in Section~\ref{sec:env}. These galaxies were selected to
be at least 2.5~mag brighter than the SDSS magnitude limit, to insure
that isolated galaxies are not ``artificially isolated'' because their
satellites lurk below the SDSS {spectroscopic sample
  threshold}. Our main conclusions 
are unchanged when considering this ``bright'' subset.

\begin{figure*}
\centering
\includegraphics[width=0.99\textwidth]{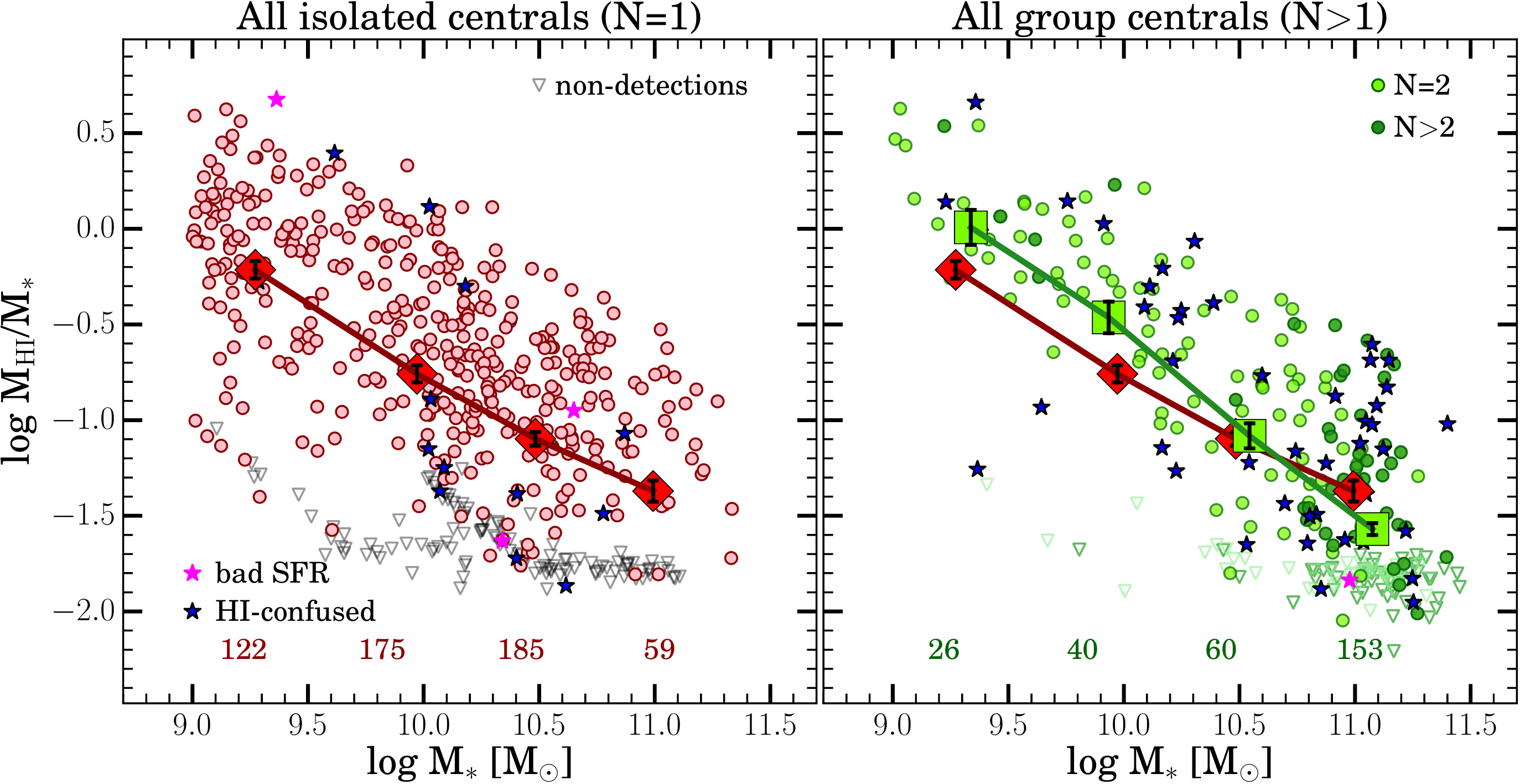}\\
\vspace{0.5cm}
\includegraphics[width=0.99\textwidth]{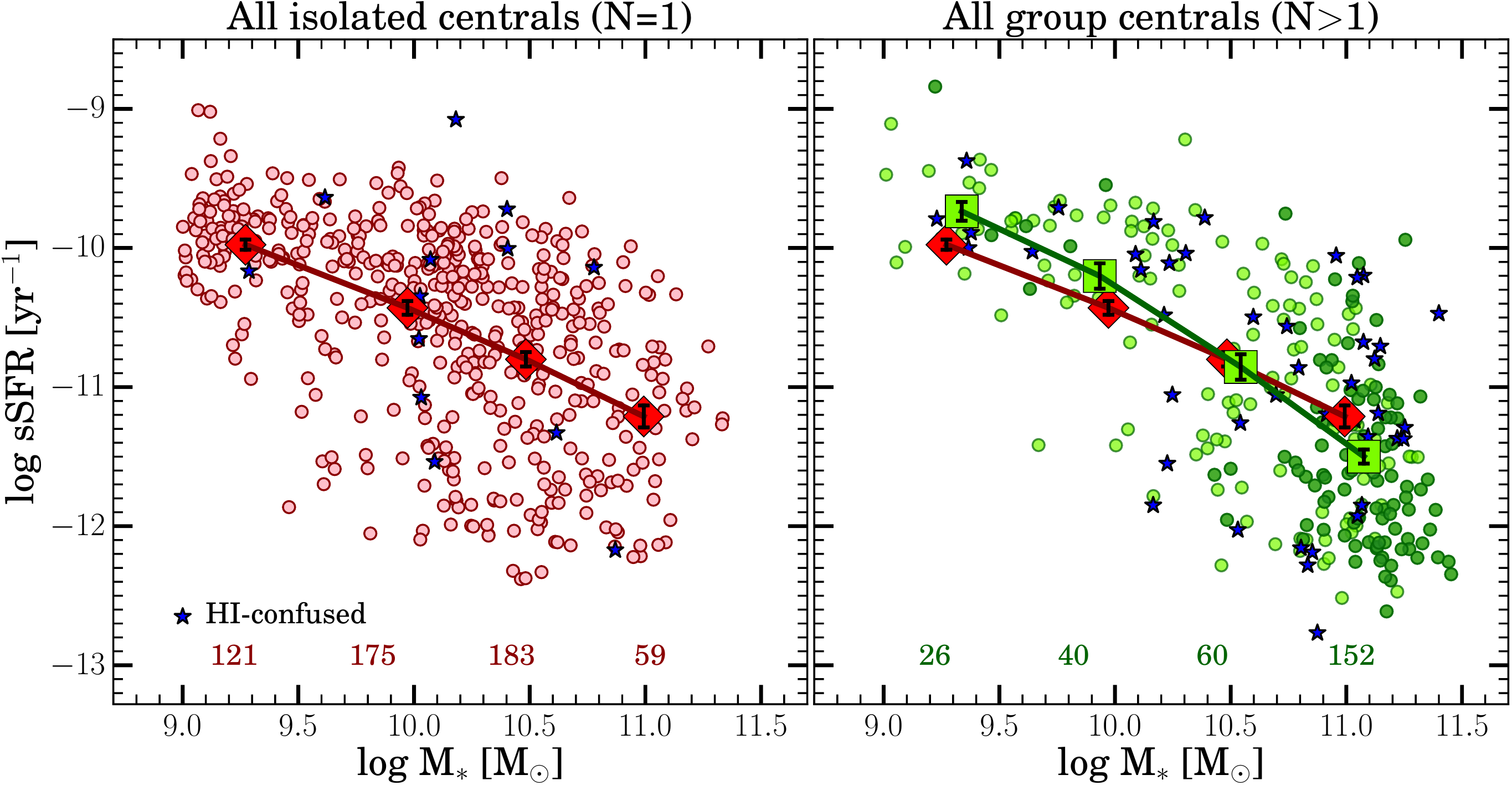}
    \caption{
      Same as Figures~\ref{fig:envHI} and \ref{fig:sfr}, but including
      all \hi-confused galaxies (i.e., galaxies whose measured \hi \,
      emission belongs completely or for the most part to another
      galaxy within the Arecibo beam) {as dark blue stars} and
      those with no SFR estimates {as magenta stars}. Our main
      results are unchanged. 
    }
    \label{fig:all_GFSF}
\end{figure*}

Figure~\ref{fig:all_GFSF} shows our main \hi \, gas fraction and sSFR
relations for central galaxies now including all \hi-confused objects
and those for which SFR estimates are unavailable. Again our results
and conclusions are not affected.

\iffalse
To test for the worst possible effects of the unobserved AA70 galaxies
(described in Section~\ref{sec:sample}), Figure~\ref{fig:GF3} shows our
\hi gas fraction relations and includes the three (un-confused) AA70
group centrals from the GASS-lo PS set to worst-case upper
limits. They are all in the second mass bin (of N=36 galaxies), and
have a small effect on its average \hi gas fraction.
\fi

\iffalse
-check adding in 4 not-observed group centrals from xGASS PS,
    see what worst-case changes happen at low limits
 these 4 are:
 GASS    ra         dec         z            lgMstar   NgB envcodeB
 109019	 140.29292   1.4727391  0.016332524  9.838872  2	 2
 113033  195.83981  28.30904    0.01842575   9.73217   2         2
 113009  198.0537    2.1209269  0.019400947  9.952491  3         2
 114066  211.11417  16.303675   0.014019729 10.004926  2         2
 --none are shredded, which is good.
 --114066 would be confused - massive interaction, so don't worry
 include 109019, 113033, 113009 on GF plot, see what happens?
-small effect on 2nd bin only. didn't check KS changes
(in appendix now)

\begin{figure*}
\centering
\includegraphics[width=0.8\textwidth]{GFM_new_with3-crop.pdf}\\
    \caption{
    Same as Figure~\ref{fig:envHI}, but including worst-case estimated
    upper limits 
    of the three unobserved AA70 group centrals. Our
      main results are unchanged.
    }
    \label{fig:GF3}
\end{figure*}
\fi

\begin{figure*}
\centering
\hspace{0.07\columnwidth}
\citet{yang07} DR7,  ``A'' catalog:  \hspace{0.075\columnwidth}
\citet{yang07} DR7,  ``B'' catalog:  \hspace{0.075\columnwidth}
\citet{yang07} DR7,  ``C'' catalog: \hspace{0.01\columnwidth} \, \\
\includegraphics[width=0.363\textwidth]{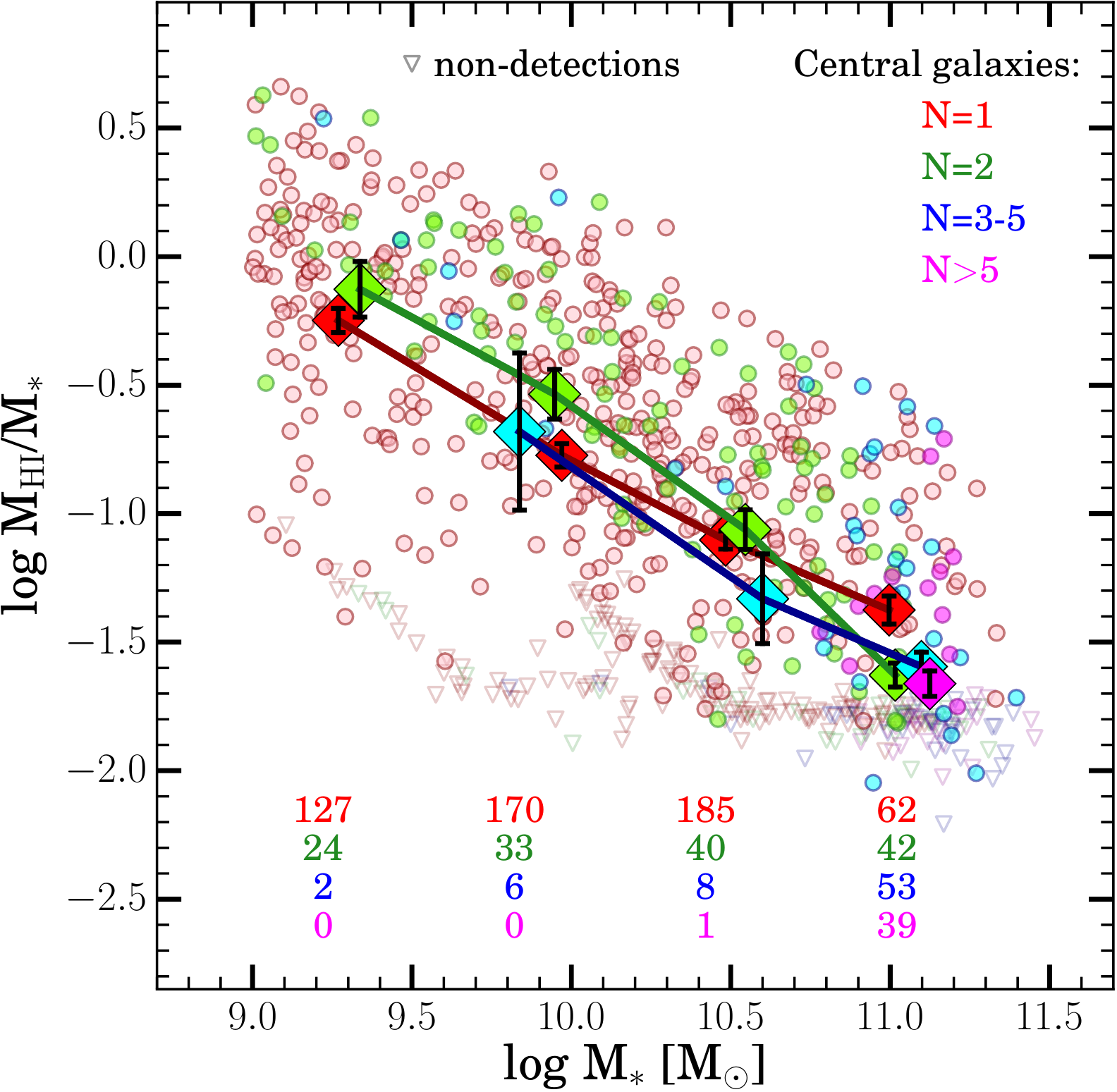}
\includegraphics[width=0.313\textwidth]{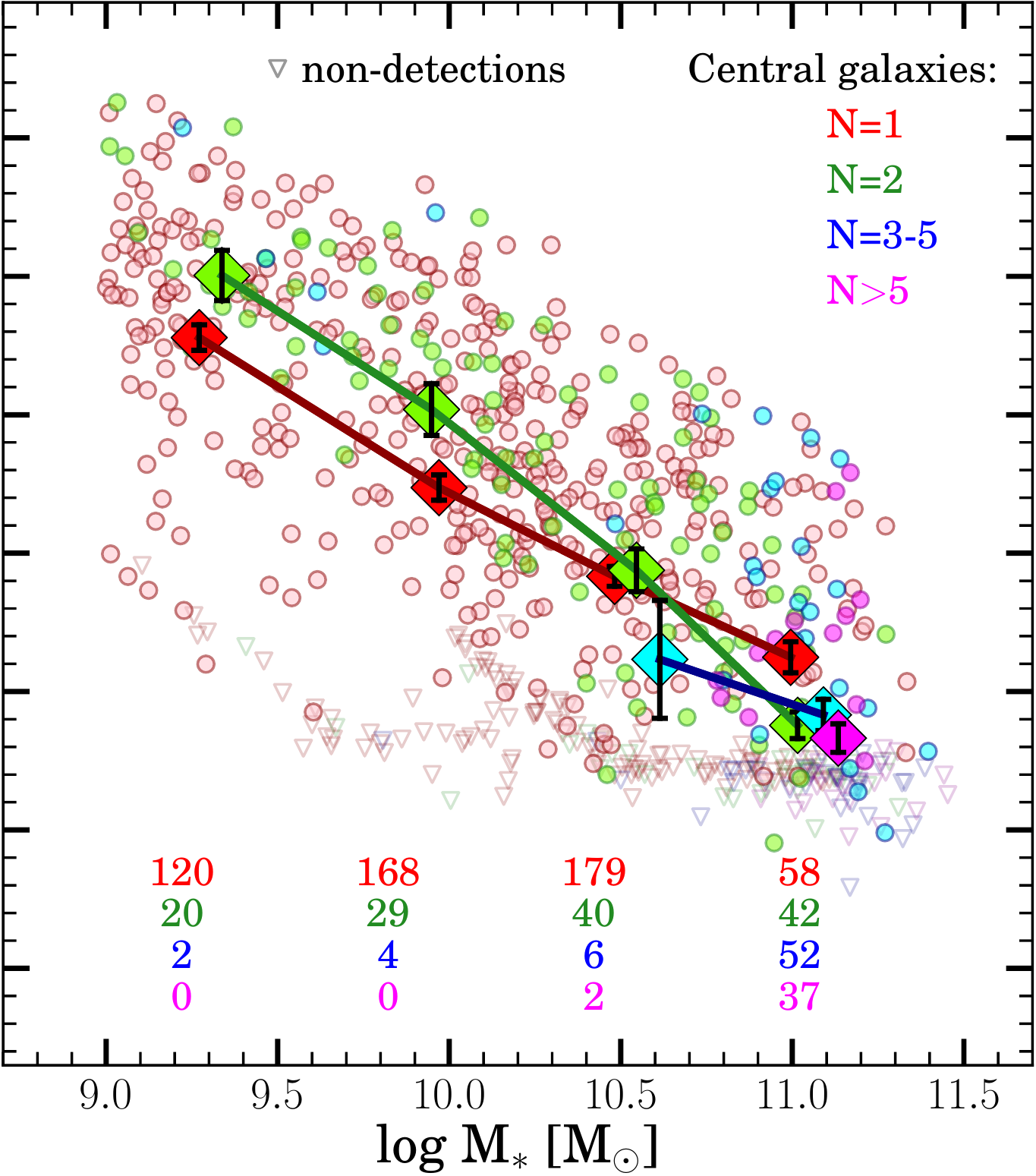}
\includegraphics[width=0.313\textwidth]{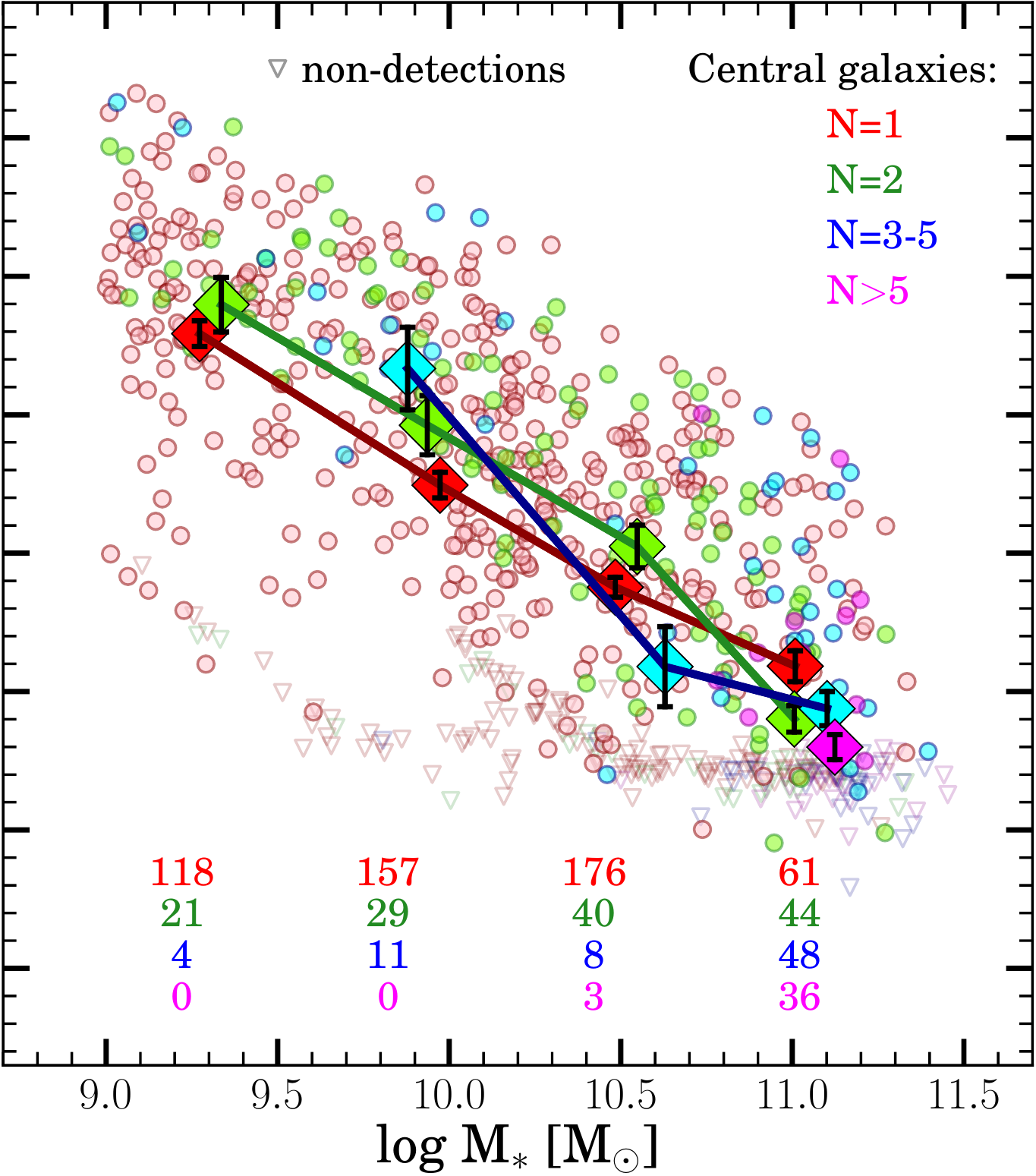}\\
    \caption{
   Same as Figure~\ref{fig:envHI}, now 
   using all three group catalogs from \citet{yang07}. Here (and on
   Figure~\ref{fig:allgroups_sfr})
   we plot isolated central {galaxies in isolation
   (N=1, red points) those in groups of different sizes, from N=2
     (green), to N=3-5 (blue), and N$\ge$5 (magenta). The elevated \hi
     \, gas fractions 
   at low stellar mass in group central galaxies are driven by the
   dominant population of N=2 groups.} No 
   significant differences between  catalogs are present, and our
   main observational results are robust 
   across group catalogs, {and when considering groups of
     different multiplicities}.
    }
    \label{fig:allgroups_GF}
\end{figure*}

\begin{figure*}
\centering
\hspace{0.07\columnwidth}
\citet{yang07} DR7,  ``A'' catalog:  \hspace{0.075\columnwidth}
\citet{yang07} DR7,  ``B'' catalog:  \hspace{0.075\columnwidth}
\citet{yang07} DR7,  ``C'' catalog: \hspace{0.01\columnwidth} \, \\
\includegraphics[width=0.363\textwidth]{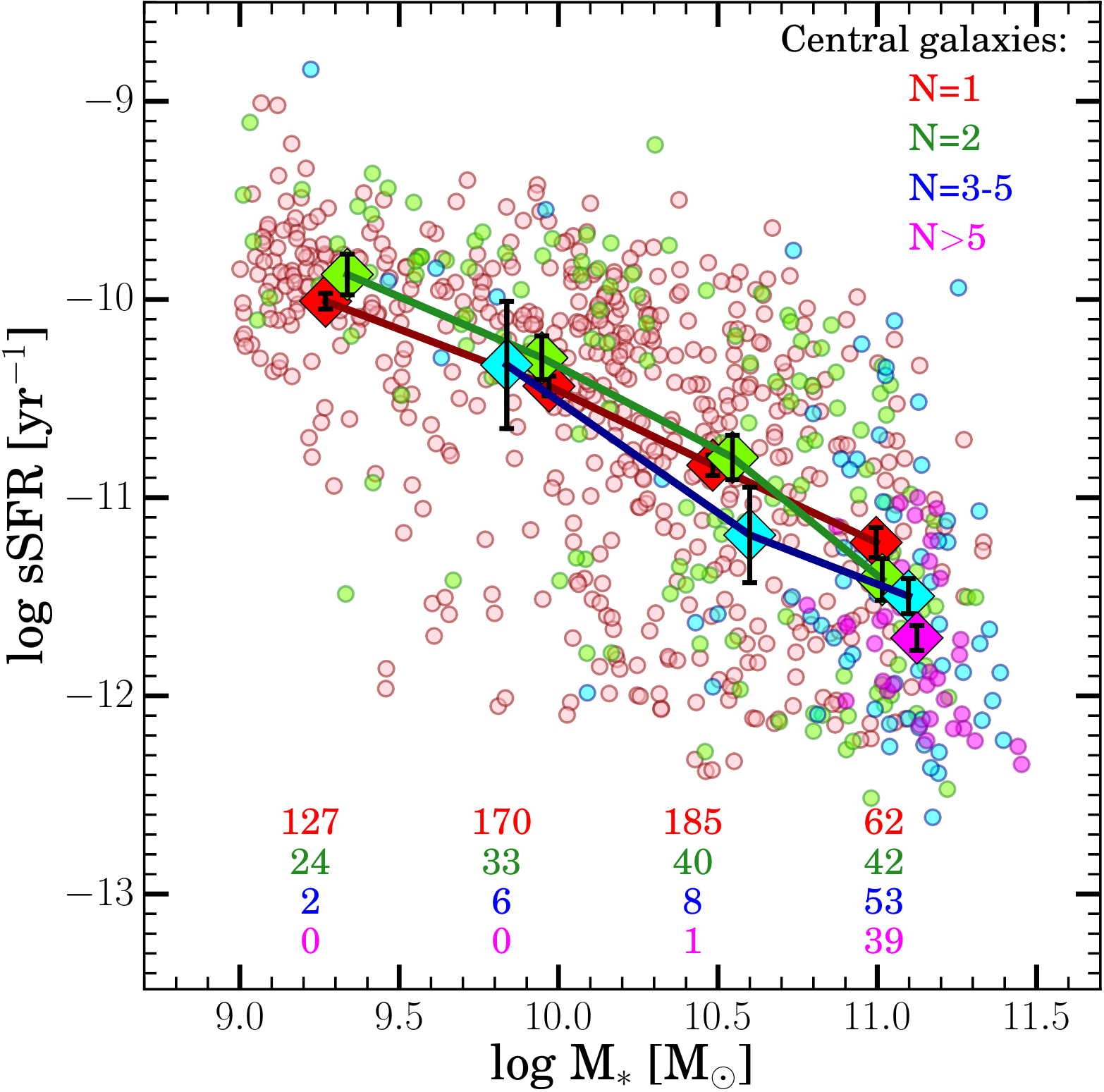}
\includegraphics[width=0.313\textwidth]{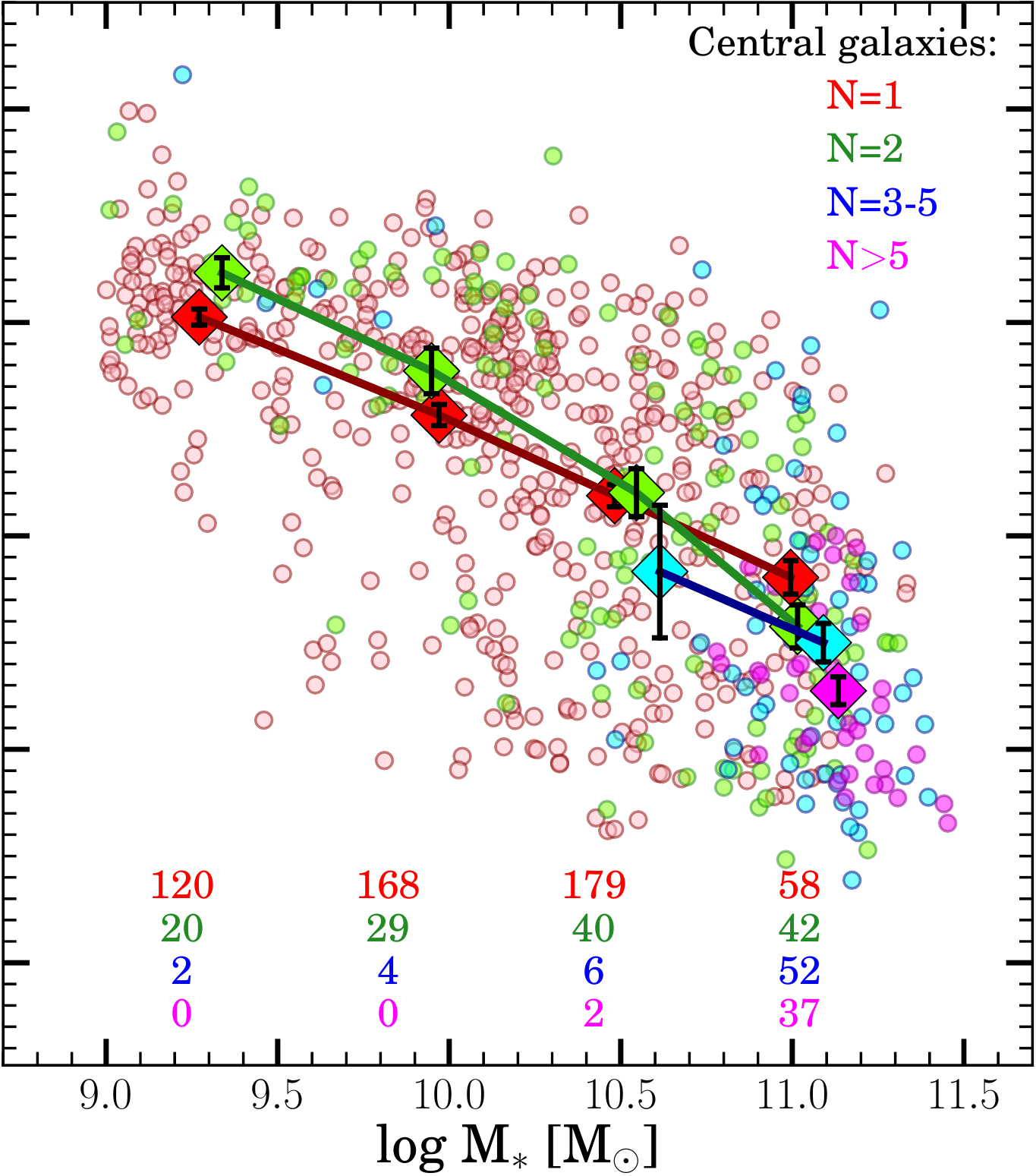}
\includegraphics[width=0.313\textwidth]{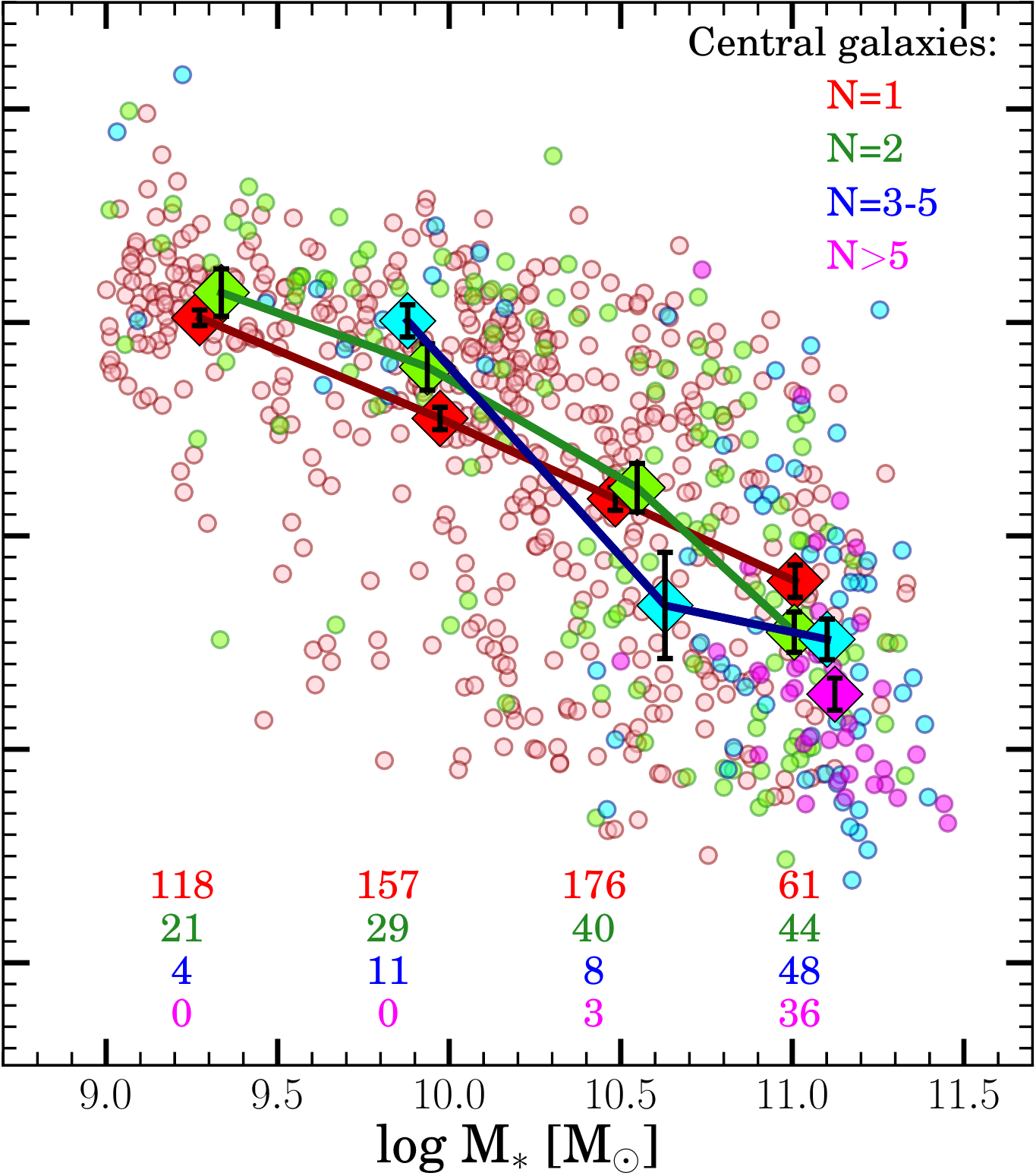}\\
    \caption{
   Same as Figure~\ref{fig:sfr}, but 
   using all three group catalogs. No significant differences between group
   catalogs are present, and our main observational results are robust
   across group catalogs.
    }
    \label{fig:allgroups_sfr}
\end{figure*}

We have also compared our group identities and group sizes with the
other versions of the DR7 group catalogs of \citet{yang07}. As
mentioned in Section~\ref{sec:env}, these versions are built using
different numbers of objects with redshifts from a variety of
sources. We consistently find that low mass group central galaxies
have larger average \hi \, gas fractions and sSFRs compared with
isolated centrals at the same stellar
mass. Figures~\ref{fig:allgroups_GF} and \ref{fig:allgroups_sfr} show
re-creations of our key results from Figures~\ref{fig:envHI} and
\ref{fig:sfr} using different group catalogs to obtain the
environmental identities of our galaxy sample. Included are the
identities from DR7 Group ``A'', ``B'', and ``C'' catalogs of
\citet{yang07}. Each version includes increasingly more galaxies (see
Section~\ref{sec:env}). We also use {finer} environmental
categories in these figures: instead of contrasting isolated central
galaxies (N=1) from group central galaxies ({N>1}), here we
{separately show group central galaxies in groups of different}
sizes. As noted in Section~\ref{sec:env}, all central galaxies in
groups with N>4 have \Mst/\msun>$10^{10.8}$.

Regardless of the group catalog used, we consistently see that at low
stellar mass (\Mst/\msun<$10^{10.2}$), group central
galaxies have elevated \hi \, gas fractions and sSFRs compared with
isolated central galaxies, {and that this difference is largely
  driven by central galaxies in groups of N=2.}

%dropped tempel/saulder since they find bigger groups and it is more
%  complicated than we want to get in to here

%\textbf{Could comment further on variations: Saulder+ actually has
%  larger groups at these low masses, and those are \emph{even more}
%  gas rich. This is consistent with our main conclusions, but goes
%  beyond groups of N=2-4. Yang A \& C catalogs have weaker trend, but
%  there are reasons to worry about those catalogs. Probably no further
%  comments are necessary about these alternative catalogs.}

% Don't change these lines
\bsp	% typesetting comment
\label{lastpage}
\end{document}

% End of mnras_template.tex